\documentclass[12pt,preprint]{aastex}

\usepackage{graphicx}
\usepackage{amsmath}
\usepackage{natbib}
\usepackage{color}
\shorttitle{Habitable Zone around Black Holes}
\shortauthors{Schnittman}

\begin{document}

\title{Life on Miller's Planet: The Habitable Zone Around Supermassive
  Black Holes} 

\author{Jeremy D.\ Schnittman}
\affil{NASA Goddard Space Flight Center, Greenbelt, MD 20771}
\affil{Joint Space-Science Institute, College Park, MD 20742}
\email{jeremy.schnittman@nasa.gov}

\begin{abstract}
In the science fiction film {\it Interstellar}, a band of intrepid
astronauts sets out to explore a system of planets orbiting
a supermassive black hole, searching for a world that
may be conducive to hosting human life. While the film legitimately
boasts a relatively high level of scientific accuracy, it is still
restricted by Hollywood sensitivities and limitations. In this paper,
we discuss a number of additional astrophysical effects that may be
important in determining the (un)inhabitable environment of a planet
orbiting close to a giant, accreting black hole. Foremost among these
effects is the blueshift and beaming of incident radiation on the
planet, due to the time dilation of an observer orbiting very close to
the black hole. This results in high-energy flux incoming from
surrounding stars and background radiation, with significant
implications for habitability.
\end{abstract}


\section{INTRODUCTION}\label{section:intro}

A favorite question in the popular discussion of black holes is,
``what would happen if the sun were to suddenly turn into a black
hole?'' On the up side, we know that the Earth would not get
``sucked in.'' Far from a black hole (and the Earth is millions of
Schwarzschild radii away from the Sun), gravity looks almost exactly
Newtonian. On the down side, the Sun provides almost all the energy
necessary for life on Earth to survive. Without it's constant heat
flux, the oceans would likely freeze over in a matter of days. 

But we also know that many astrophysical black holes can provide their
own energy source, in the form of radiation from hot, accreting
gas. In fact, for most observable black holes, this accretion power
outweighs anything attainable from nuclear fusion by many orders of
magnitude. So one could naturally imagine that replacing the Sun with
an accreting black hole might not be the end of life on earth after
all. 

This entertaining thought experiment is not unlike the premise of the
science fiction blockbuster {\it Interstellar}. The only
difference is that, instead of replacing the Sun with a black hole,
due to an imminent biological collapse on Earth, humanity is forced to
leave the Solar System behind and seek out a new habitable
world orbiting a distant black hole. \textcolor{red}{[Warning! Spoiler
    alert!]} Advanced probes have identified three potentially
habitable planets orbiting a supermassive black hole called {\it
  Gargantua}. NASA then sends a follow-up team of astronauts and
scientists to explore these three targets in detail, looking for more
promising evidence for habitability.

In this paper, we ask the simple question: what could we know {\it a
  priori} about the environments of these planets, and in particular,
the prospect of their habitability? On the face of it, this is nothing
more than the typical armchair speculation and commentary disseminated
by generations of sci-fi fans poking their fingers into any perceived
plot hole. Yet we hope to show that the question of habitability
around supermassive black hole is an extremely valuable pedagogical
tool. Among the important physics problems involved are accretion
dynamics, general relativity, tidal evolution, atmospheric chemistry,
and astrobiology. 

Many of these questions are already asked and answered in the movie's
excellent companion book {\it The Science of Interstellar}
\citep{Thorne:2014}, a source on which we will naturally lean
heavily. The primary novelty of this paper is the expanded discussion
of habitability for exoplanets in general, along with a focus on the
important implications of the extreme time dilation found on Miller's
planet. 

It cannot be overemphasized that the purpose of this paper is {\it
  not} a critique of {\it Interstellar}, neither from the point of
view of film quality or scientific accuracy. Rather, it should be
understood as a form of fan fiction, exploring in greater detail and
new directions the world created by Christopher Nolan and Kip Thorne.

\section{DEFINITION OF HABITABILITY}\label{section:habitability}
We begin with the simple question of what makes a planet habitable?
This is one of the most important and active areas of research in the
exoplanet field today. Much like the multi-step triage strategy
employed by the explorers in {\it Interstellar}, the NASA roadmap for
exoplanet exploration involves a multi-stage approach where smaller
probe missions (e.g., TESS, the Transiting Exoplanet Survey Satellite;
\citet{TESS2015})
are used to identify promising targets for intense spectroscopic
observations with the James Webb Space Telescope. Eventually, the goal
is to find promising targets for direct imaging of Earth-like planets
by future flagship missions such as the Large-aperture UV/O/IR
Telescope.

These future direct imaging missions will ultimately have the goal of
detecting biosignatures by identifying specific spectral features in
the planet's atmosphere. In the most plausible scenarios (in our
admittedly limited, Earth-centric imagination), atmospheric
biosignatures require the existence of liquid water on the planet's
surface. Thus, to first order, the {\it habitability zone} (HZ) for a
planetary system is defined as the region where the equilibrium
temperature on the planet's surface is between 273 and 373 K. 

To determine a planet's equilibrium surface temperature, one simply
balances the incoming heat sources with the outgoing radiative
flux. Of course, in practice, this calculation can be anything but
simple. For example, the total Solar flux incident on the Earth is
1250 W/m$^2$. This flux is incident on a surface with cross sectional
area $\pi R_E^2$ yet the radiated flux is emitted from an area of
$4\pi R_E^2$. Assuming a blackbody emissivity law, the effective
temperature can be solved from
\begin{equation}\label{eqn:Teff}
4\pi R_E^2 \sigma T_{\rm eff}^4 = \pi R_E^2 F_\odot \, ,
\end{equation}
giving $T_{\rm eff} = 277$ K. The actual value, averaged over the Earth's
surface, is $T_{\rm eff} = 288$ K. 

It turns out that this close agreement is something of a happy
coincidence, since we have ignored several important physical
processes. First, not all of the solar radiation is actually absorbed
by the Earth. Some is reflected off of clouds, icecaps, and even the
ocean. This reflection is quantified by the {\it albedo} $\alpha$,
defined as the ratio of the reflected to incident flux. For the Earth,
$\alpha \approx 0.3$, meaning only $70\%$ of the incident solar radiation
actually contributes to warming the surface. So we should simply modify
equation (\ref{eqn:Teff}) by multiplying the right-hand-side by
$(1-\alpha)$, giving an equilibrium temperature of 254 K, nearly 20
degrees below freezing!

However, the left-hand-side of (\ref{eqn:Teff}) is also wrong, since
not all the radiation emitted from the surface actually escapes to
space. While the atmosphere is largely transparent to the incoming
solar flux, which peaks at wavelengths around 400--600 nm, the
outgoing infrared radiation ($\sim$10-15 $\mu$m) is much easier to
absorb in the atmosphere, specifically by molecules like H2O, CO2, and
CH4. This is the well-know ``greenhouse effect,'' which for Earth
leads to a net warming of $\sim $34 K over the simple equilibrium
temperature, and gives us a warm, stable planet on which life can
thrive.

What about other planets? Considering only solar system objects, the
range of albedos on rocky bodies ranges from 0.1 for Mercury to 0.7
for cloud-covered Venus. Furthermore,
the magnitude of the greenhouse effect quoted above is only applicable
for Earth's present atmosphere and the Sun's present
spectrum. Changing either significantly could lead to vastly different
figures. One need only look at Earth's twin Venus to appreciate the
potential implications of greenhouse warming. Largely motivated by
Venus's radically different climate, \citet{Kasting1993} employed a
vertical 1D climate calculation with a detailed photochemical code so
as to better understand the processes that could lead to a runaway
greenhouse effect. They found that, as the incident solar flux
or the mixing fraction of CO2 increased above a critical
value, the surface temperature would increase to the point where water
vapor (a powerful greenhouse gas) would begin to dominate the
atmosphere, further heating the planet until the oceans evaporated
entirely. Once in the upper atmosphere, the H2O could dissociate and
the light H2 could escape entirely, leaving a hot, dry, CO2-dominated
planet behind. Without liquid water, it becomes very difficult for the
atmospheric CO2 to get re-captured through weathering processes common
on Earth, and thus accumulates in the atmosphere. 

\begin{figure}[h]
\caption{\label{fig:HZ} Habitable zone for planets with Earth-like
  atmospheres around a variety of stellar types. [reproduced with
    permission from \citet{Kopparapu2013}]}
\begin{center}
\scalebox{0.6}{\includegraphics{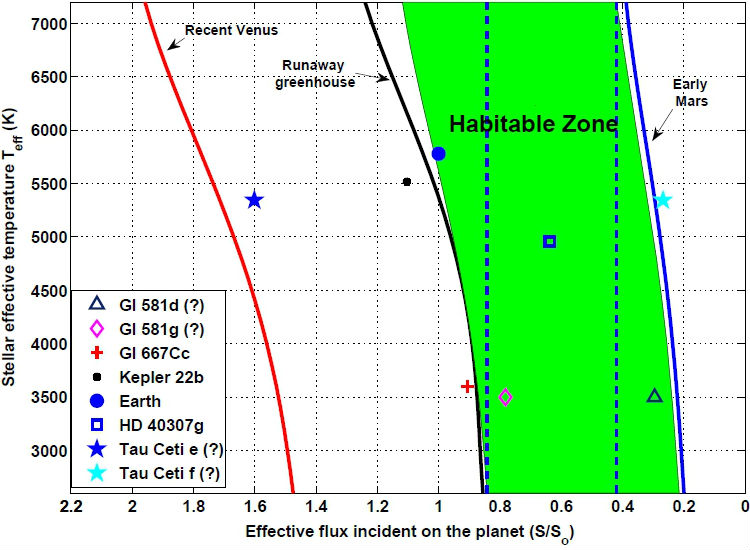}}
\end{center}
\end{figure}

In the two decades since these early calculations, the 1D coupled
climate-photochemistry codes have evolved significantly, incorporating
more detailed opacity models as well as a more diverse range of
chemical composition for the atmosphere. The current state-of-the-art
results for the habitable zone can be found in \citet{Kopparapu2013},
reproduced in Figure \ref{fig:HZ}. The figure shows the
``conservative'' HZ as the region between the two blue curves, and the
``optimistic'' HZ between the red and orange curves. 

Somewhat ominously, the Earth is right at the inner edge of the
conservative HZ, at clear risk of crossing the line to runaway
greenhouse. Yet the calculations in \citet{Kopparapu2013} are only
one-dimensional, and necessarily neglect many complicating factors
such as cloud formation, which we will see below may have an important
stabilizing effect on the climate. 

We can also see in Figure \ref{fig:HZ} the effects of
changing the host star's mass, and thus the incident spectrum on the
planet. Smaller stars are much less luminous, and also cooler,
producing a blackbody spectrum peaking closer to the infrared. Thus
the HZ for faint M-dwarf systems is well inside 0.1 AU. 

Also plotted (dotted line) in Figure \ref{fig:HZ} is the tidal locking
radius. Any terrestrial planet inside this point will likely be
tidally locked to the host star on the timescale of a Gyr or
less. Note that Mercury is just inside this line, consistent with the
fact that it is quasi-locked to the Sun, trapped in a 3:2 spin-orbit
resonance \citep{Murray:1999}. In the Newtonian regime, the tidal
forces are quite easy to calculate, and scale simply with the stellar
mass to the one-third power:
\begin{equation}\label{eqn:tl}
R_{\rm tl} \propto R_p (M_\ast/M_p)^{1/3} \, .
\end{equation}
For rocky planets of similar composition and thus density, $R_p \sim
M_p^{1/3}$ so the tidal locking radius $R_{\rm tl}$ is a function only
of the stellar mass $M_\ast$. 

\begin{figure}[h]
\caption{\label{fig:tl_BH} Tidal locking radius for rocky planets
  orbiting supermassive black holes as a function of black hole mass;
  the radius is measured in (a) astronomical units and (b)
  gravitational radii. Any planet inside $R_{\rm tl}$ will be tidally
  locked to the black hole.}
\begin{center}
\scalebox{0.45}{\includegraphics*{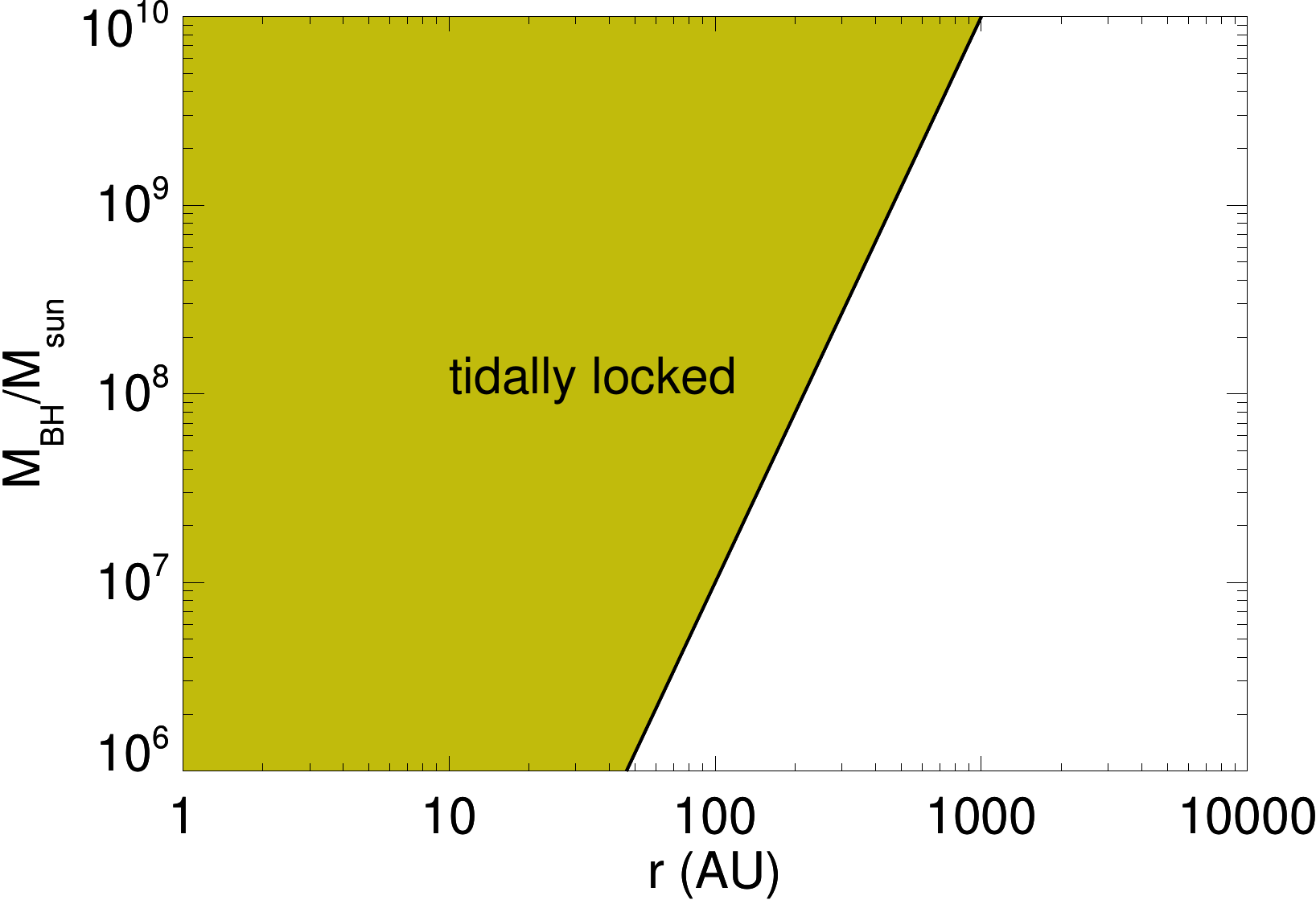}}
\scalebox{0.45}{\includegraphics*{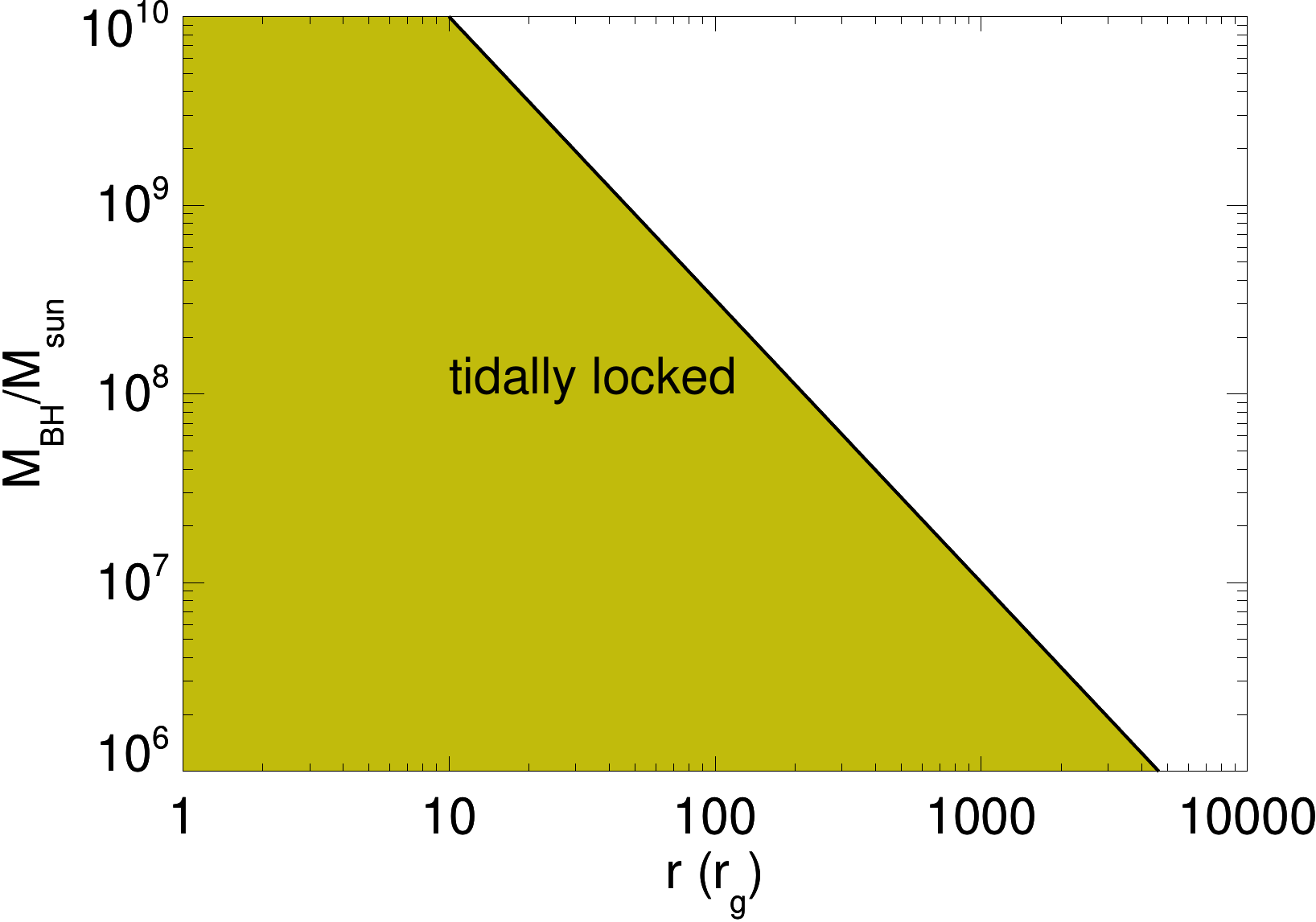}}
\end{center}
\end{figure}

Scaling equation (\ref{eqn:tl}) up to black hole masses naturally
gives much larger tidal locking radii, as shown in Figure
\ref{fig:tl_BH}a. As in many studies of black holes, it is convenient
to define the {\it gravitational radius} $r_g \equiv GM/c^2$ as a unit of
length. This provides a valuable estimate for when relativistic
effects become important, as first-order post-Newtonian corrections
are proportional to $(v/c)^2 \sim (r/r_g)^{-1}$. Thus in Figure
\ref{fig:tl_BH}b we plot the tidal locking radius as a function of
gravitational radii. This curve is determined simply by scaling up the
Newtonian relationship to black hole masses. For black holes less than
$\sim 10^9 M_\odot$, it seems likely that post-Newtonian effects will
be unimportant on determining the specific tidal-locking radius. Above
$10^9 M_\odot$, the relativistic form of the tidal tensor must be
used, e.g., with locally flat Fermi normal coordinates, as in
\citet{Cheng:2013}. 

Much closer to the black hole (or more likely, for much smaller black
holes), the gravitational forces could be so great as to tidally
disrupt the planet. The tidal disruption radius $R_{\rm td}$ scales
just like the locking radius $R_{\rm tl}$, but roughly two hundred times
smaller: $R_{\rm tl} \approx 200 R_{\rm td}$. From Figure
\ref{fig:tl_BH}, we see that a terrestrial planet would get tidally
disrupted just inside the horizon of Gargantua, making Miller's
planet deformed but not disrupted \citep{Thorne:2014}.

We can see from Figure \ref{fig:tl_BH} that for Gargantua's
mass of $10^8 M_\odot$ \citep{Thorne:2014}, any planet inside a radius
of 100 AU will be tidally locked to the black hole. Conveniently
enough, for this mass, one AU is almost exactly one $r_g$. 

For tidally locked planets, the incoming flux is always
absorbed at the same part of the planet. For planets around regular
stars, this is called the ``sub-stellar'' point, directly facing the
star. As we will see below, for planets around black holes, this will
generally be a point facing in the direction of the planet's orbit. In
both cases, the absorbed flux must be transported around the planet
via ocean or atmospheric currents, or else the dark half of the planet
will freeze over. 

\begin{figure}[h]
\caption{\label{fig:Yang} Cloud coverage maps with overlaid
  temperature contours for (left) rapidly and (right) slowly rotating
  planets. The strong cloud coverage at the sub-stellar point for
  tidally locked planets has a significant cooling effect on the
  surface temperature. [reproduced with permission from \citet{Yang:2014}].}
\begin{center}
\scalebox{0.4}{\includegraphics{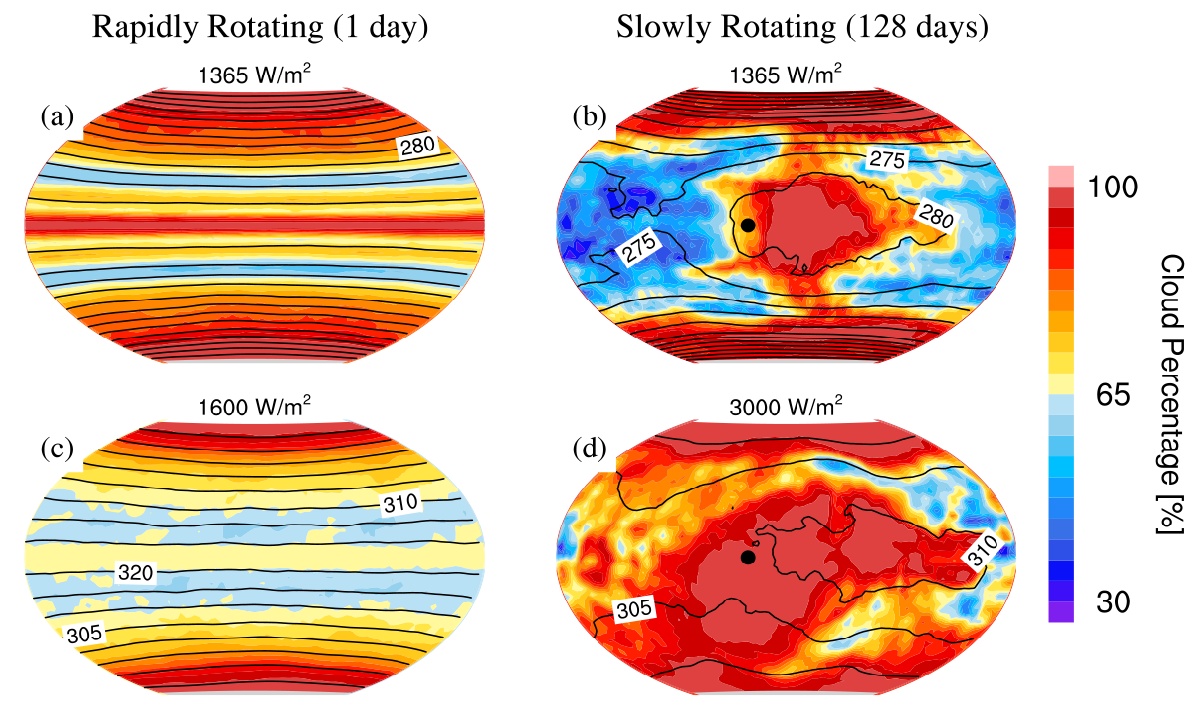}}
\end{center}
\end{figure}

Furthermore, the slow rotation of tidally locked planets will
completely change the global circulation patterns that are familiar on
rapidly-rotating planets like Earth. In particular, on Earth warm,
moist air is heated around the equator, which then rises, cools, and
forms a band of clouds and precipitation around the tropics. After
releasing the bulk of its moisture near the equator, the resulting
drier, cooler air flows towards the poles, eventually sinking in the
sub-tropical arid latitudes around $30^\circ$ (see
Fig.\ \ref{fig:Yang}a).

For tidally locked planets, instead of a band of rising air around the
equator, there is more like a single stationary chimney at the
sub-stellar point, lofting moisture in a pillar like a huge
cumulonimbus cloud. This persistent cloud coverage acts like a sun
shade, reflecting the vast majority of the incoming flux. To
investigate this effect in detail, \citet{Yang:2014} ran a full 3-D
global circulation model (GCM) of an Earth-like planet with 1-day and
128-day rotation periods for a range of incoming flux. Even for
(surface-integrated) fluxes as large as 750 W/m$^2$, the mean surface
temperatures of the slowly rotating planet was a balmy 32
$^\circ$C. Temperature contours and cloud coverage maps from their
simulations are shown in Figure \ref{fig:Yang}. 

With these GCM results in hand, we will adapt an optimistically broad
habitable zone for tidally locked planets, covering a range of
incident flux ranging from 800 W/m$^2$ at the inner edge, down to 120
W/m$^2$ at the outer edge. For planets outside the tidal locking
orbit, we will still adopt a relatively optimistic inner edge to the
HZ of 600 W/m$^2$, corresponding roughly to the curve labeled ``recent
Venus'' in Figure \ref{fig:HZ}. This is based on empirical evidence
that Venus has not had any surface water for at least the past Gyr
\citep{Solomon:1991}, so any Earth-like planet with comparable flux is
likely uninhabitable.
We will also assume that the energy balance
is due only to bolometric flux. In other words, we assume the
atmosphere is able to reprocess incoming flux over a wide range of
wavelengths, although this will also certainly break down at a certain
point. 

\section{Energy Sources}\label{section:sources}

Now that we have defined the habitable zone in terms of a planet's
tidal properties and incident flux, the next step is to explore the
range of potential energy sources for planets around supermassive
black holes. 

\subsection{Accretion Disks}\label{section:disks}

Most of what we know about black holes comes from observing the
electromagnetic radiation coming from gas as it accretes onto the
black hole. Accreting stellar-mass black holes are the brightest X-ray
sources in the sky, and accreting supermassive black holes are the
most luminous persistent sources in the universe. When the accretion
rate is sufficiently high (probably above $\sim 10^{-2} \dot{M}_{\rm
  Edd}$, with the Eddington rate defined as $\dot{M}_{\rm Edd}\equiv L_{\rm
  Edd}/(\eta c^2) \approx 10^{26} (M/10^8 M_\odot)$ g/s), the gas
likely forms an optically thick, geometrically thin accretion disk. 

Considering the motivation behind this paper, it only seems fitting to
use the \citet{Novikov:1973} model for accretion onto a Kerr black
hole. Recent, more physical models based on magneto-hydrodynamic
simulations differ mostly in the plunging region around black holes
with low to moderate spin parameters \citep{Schnittman:2016}. For
Gargantua, with a spin of $a/M=1-10^{-14}$, Novikov-Thorne should be
quite sufficient. In the Novikov-Thorne accretion disk model, the
gas moves on equatorial, circular geodesic orbits outside of the
inner-most stable circular orbit (ISCO), and then plunges rapidly into
the horizon. The local temperature of the gas increases with
decreasing radius until it reaches a maximum a few $r_g$ outside the
ISCO, and then drops to zero at the ISCO. 

\begin{figure}[h]
\caption{\label{fig:T_NT} Local fluid temperature of Novikov-Thorne
  accretion disk around a $10^8 M_\odot$ black hole accreting at
  $0.1\dot{M}_{\rm Edd}$. }
\begin{center}
\scalebox{0.8}{\includegraphics{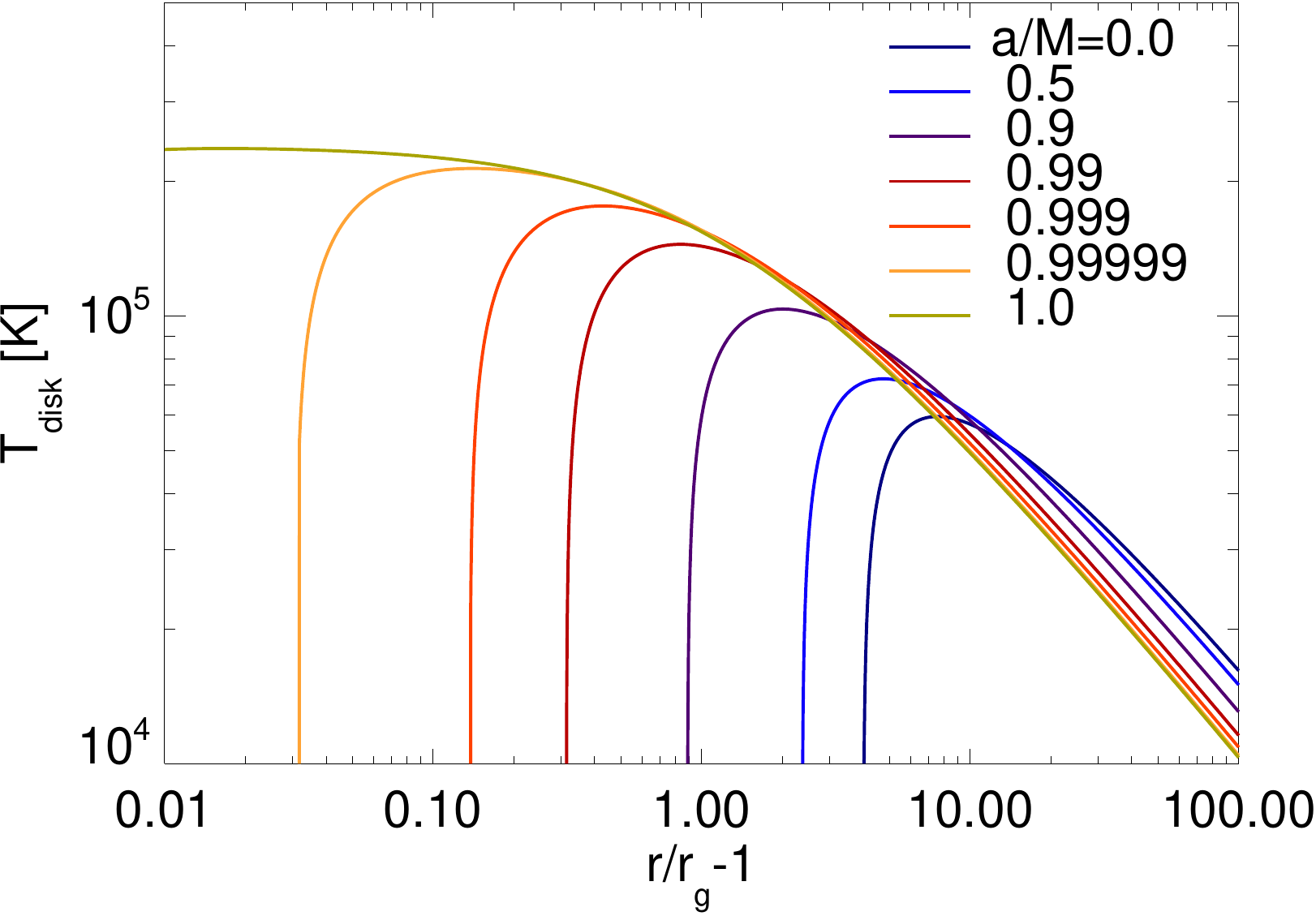}}
\end{center}
\end{figure}

In Figure \ref{fig:T_NT} we
plot the accretion disk temperature for a $10^8 M_\odot$ black hole
accreting at 10\% of the Eddington rate, for a range of spin
parameters. Note the somewhat unusual radial coordinates $r/r_g-1$,
which allow us to zoom in on the near-horizon region for nearly
maximally spinning black holes. For these nominal parameters, we see
that the peak temperature is around $10^5$ K, consistent with the fact
that most quasar spectra peak in the ultraviolet. The temperature
scales with black hole mass and accretion rate like 
\begin{equation}\label{eqn:T_peak}
T_{\rm peak} \approx 2\times 10^5 \left(\frac{M}{10^8
  M_\odot}\right)^{-1/2} \left(\frac{\dot{M}}{0.1 \dot{M}_{\rm
    Edd}}\right)^{1/4}\mbox{ K}\, .
\end{equation}

So if we want the
accretion disk to look more like a main sequence star (indeed, the
visualization of Gargantua's accretion disk does appear a very similar
color to our own Sun), we need to scale down the accretion rate by a
factor of a million. Yet even after doing this, Miller's planet,
orbiting just outside the horizon, will be completely surrounded by a
6000-degree blackbody radiation field: hardly hospitable to life! 

\begin{figure}[h]
\caption{\label{fig:disk_embed} Habitable zone for planets embedded in
  a thin accretion disk around a Kerr black hole with $a/M=1$ and mass
  $10^8 M_\odot$ (solid lines, yellow shading) and $10^9 M_\odot$
  (dashed lines, red shading).}
\begin{center}
\scalebox{0.8}{\includegraphics{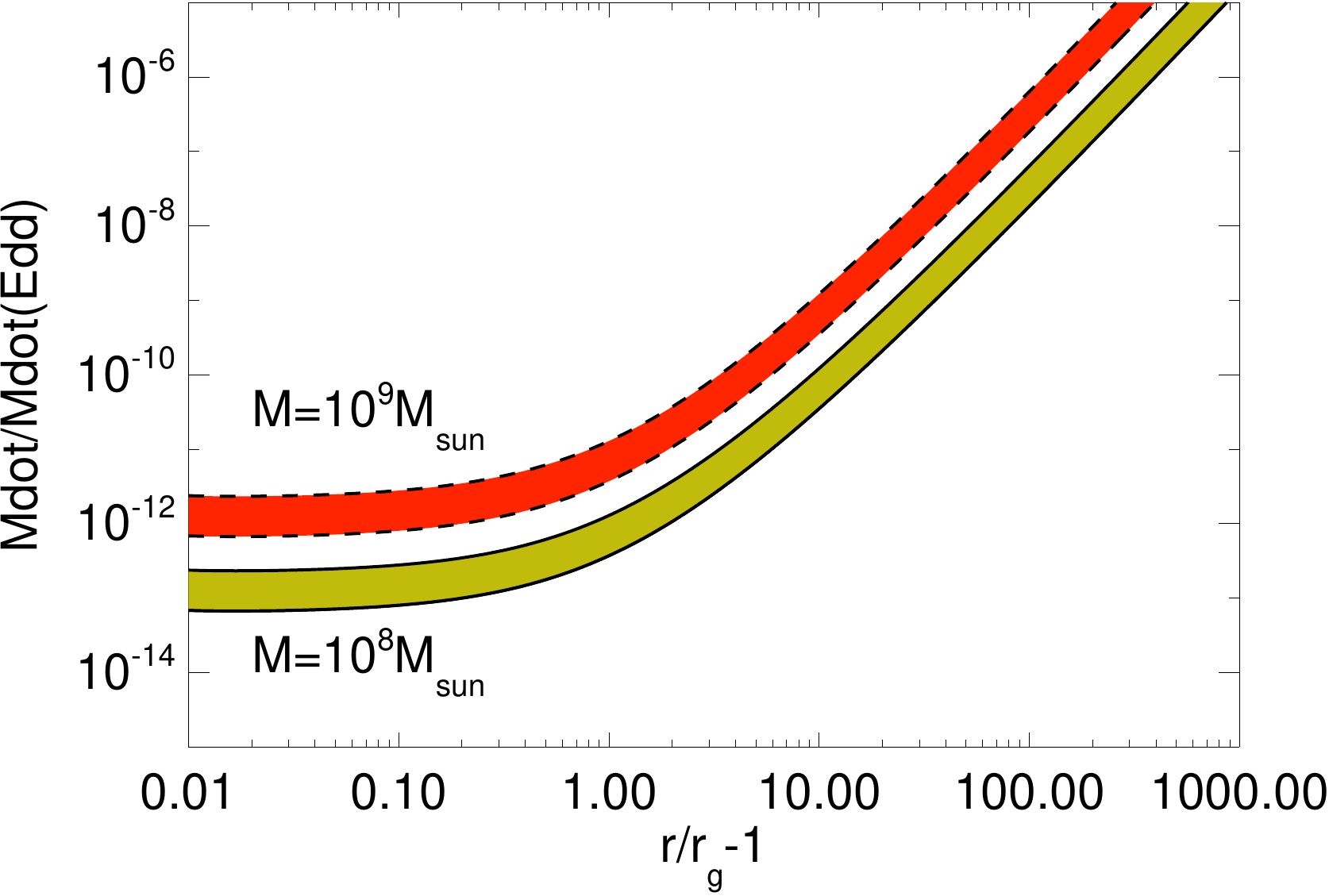}}
\end{center}
\end{figure}

As seen in Figure
\ref{fig:T_NT}, the disk temperature falls off with distance from the
black hole, roughly as $T(r) \sim r^{-3/4}$. Thus a planet located at
$\sim 100 r_g$ should be immersed in a much more comfortable
room-temperature bath of warm gas. This can be seen in Figure
\ref{fig:disk_embed}, where we plot the extent of the HZ on the x-axis
for a given accretion rate on the y-axis (alternatively, we can read
this plot as saying, for a given semi-major axis, what range of
accretion rates allow for habitability). In this scenario, we are
using the HZ flux range of 120-600 W/m$^2$ even for the tidally locked
regime, because a planet embedded in an optically thick accretion disk
has not preferred direction from which the flux is greater, and thus
dense cloud coverage will not be relevant to protecting the planet
from stellar irradiation.

There are a couple problems with the results of Figure
\ref{fig:disk_embed}. First of all, if you were to immerse Earth in a
bath of warm radiation at a temperature of 300 K, it would certainly
be habitable according to our simple requirement of liquid water on
the surface. However, all known life forms require an energy
\emph{gradient} in order to survive, so an all-pervasive blackbody
radiation background would probably not be very conducive to complex
life. Certainly not photosynthesis, which requires photon energies
sufficient to break key molecular bonds.

The other, admittedly minor, problem with the embedded disk scenario is
that, as clearly seen in the movie \emph{Interstellar}, Miller's
planet is located outside the accretion disk. This is somewhat
surprising, and would be dynamically unlikely unless there were a
sufficient gap in the disk, and the planet were endowed with some
non-zero orbital inclination. Alternatively, we could imagine the
accretion disk so geometrically thin that it just barely covered the
planet's equator, with the rest of the planet poking out above and
below the disk. Indeed, according to the classical Novikov-Thorne
model \citep{Novikov:1973}, for the black hole masses accretion rates
covered in Figure \ref{fig:disk_embed}, the disk's thickness would be
on the order of a kilometer or less. 

\begin{figure}[h]
\caption{\label{fig:diskfig} All-sky view of a thin accretion disk
  from an observer just above or below the disk. The black hole has
  spin $a/M=1$ and the accretion disk extends from $r=1M-20M$ with a
  Novikov-Thorne emissivity profile. As the observer moves closer to
  the black hole, the disk and the horizon both fill a greater fraction of the sky, and
  multiple images are apparent, due to photons circling the black hole
  multiple times before reaching the observer.}
\begin{center}
\scalebox{0.5}{\includegraphics{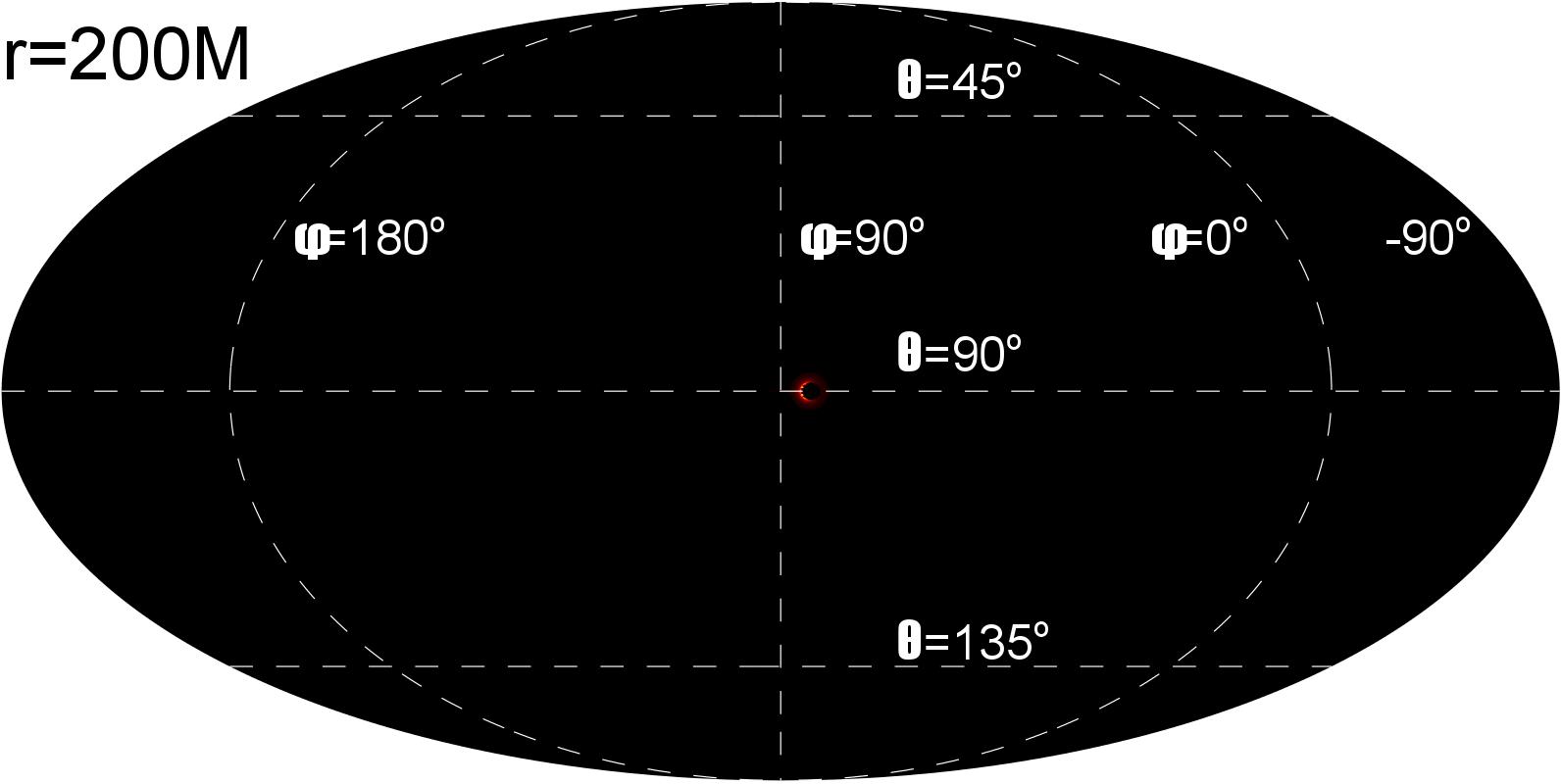}}\
\hspace{0.1cm}
\scalebox{0.5}{\includegraphics{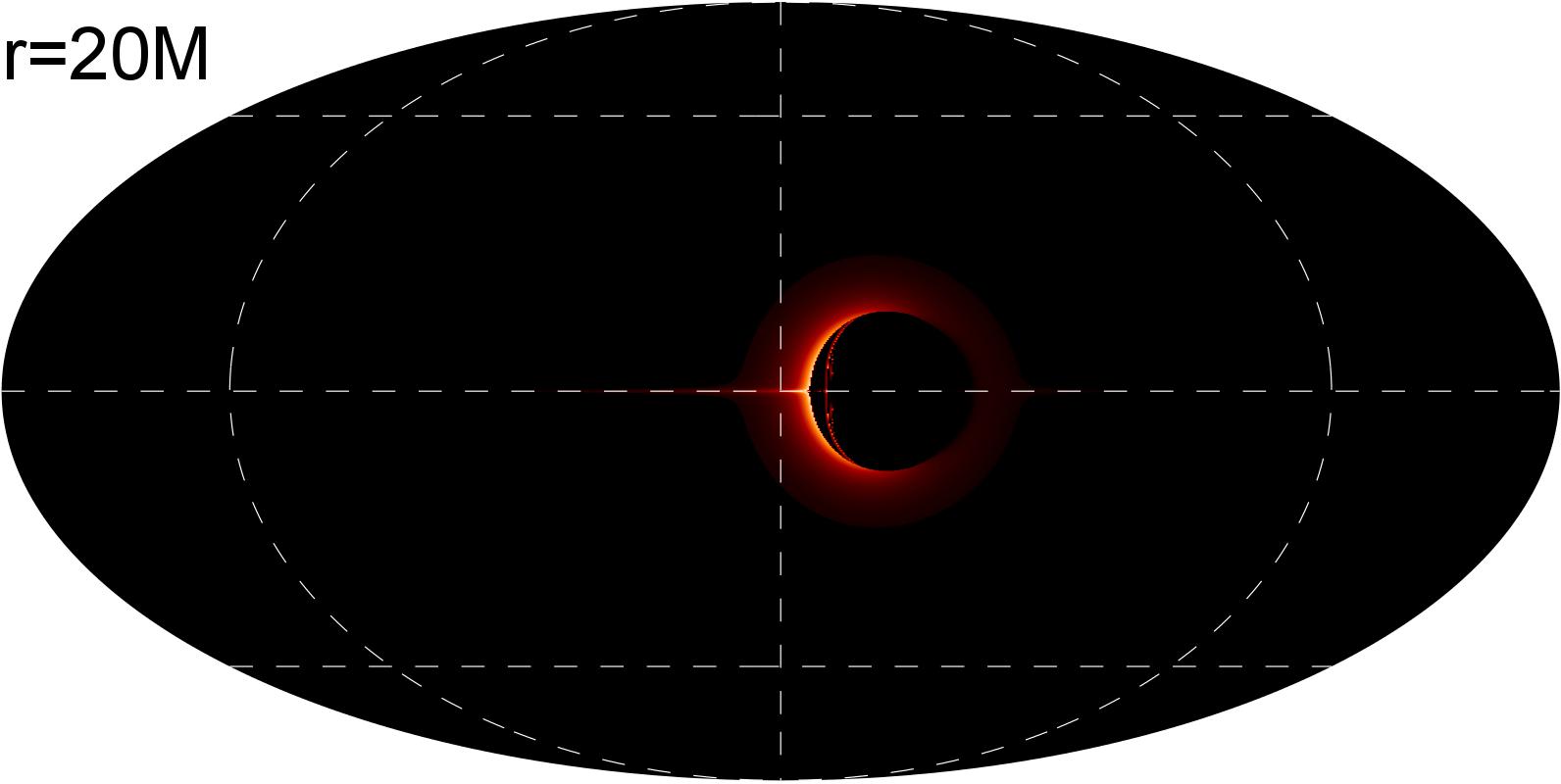}}\\
\scalebox{0.5}{\includegraphics{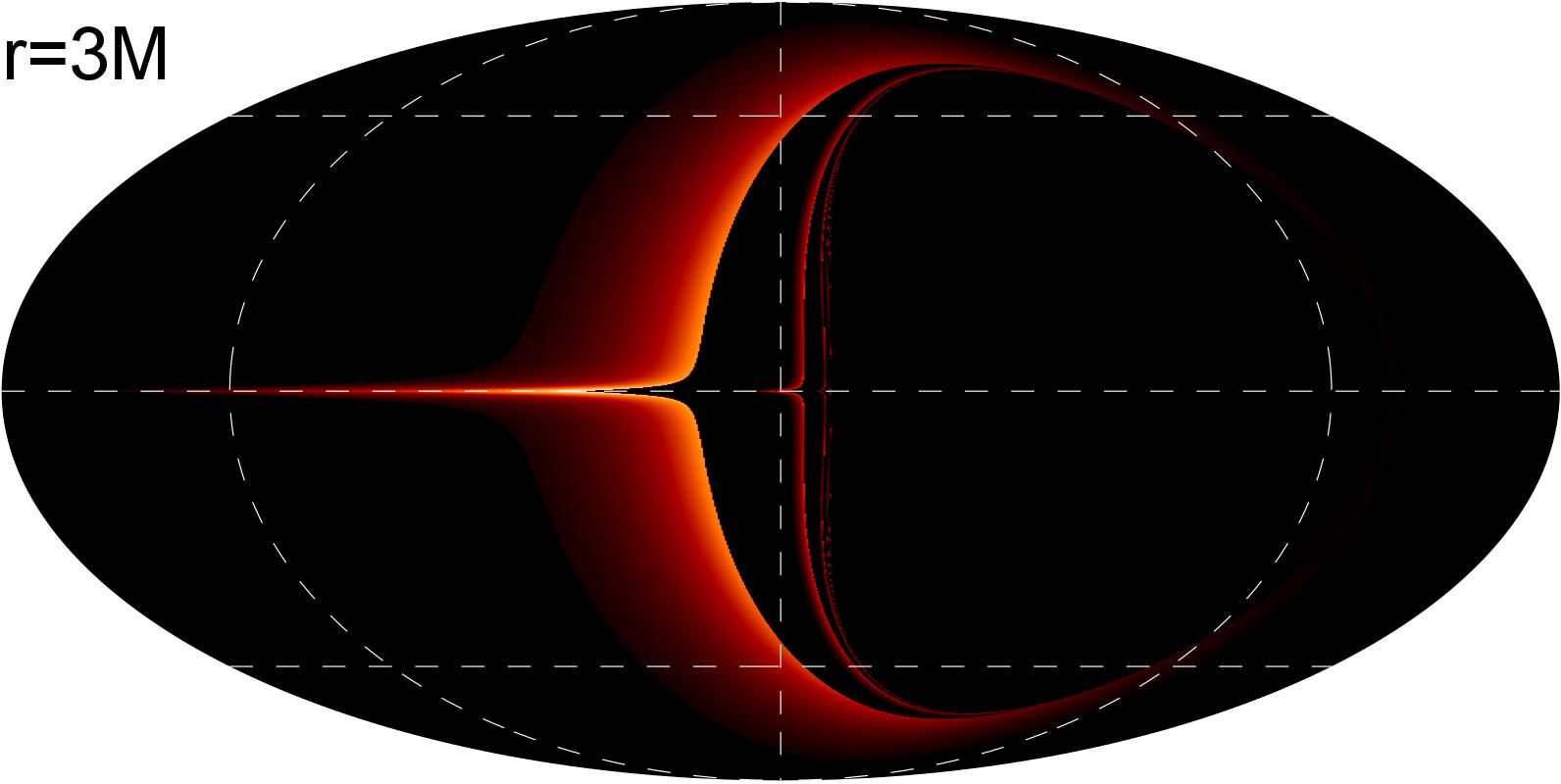}}
\hspace{0.1cm}
\scalebox{0.5}{\includegraphics{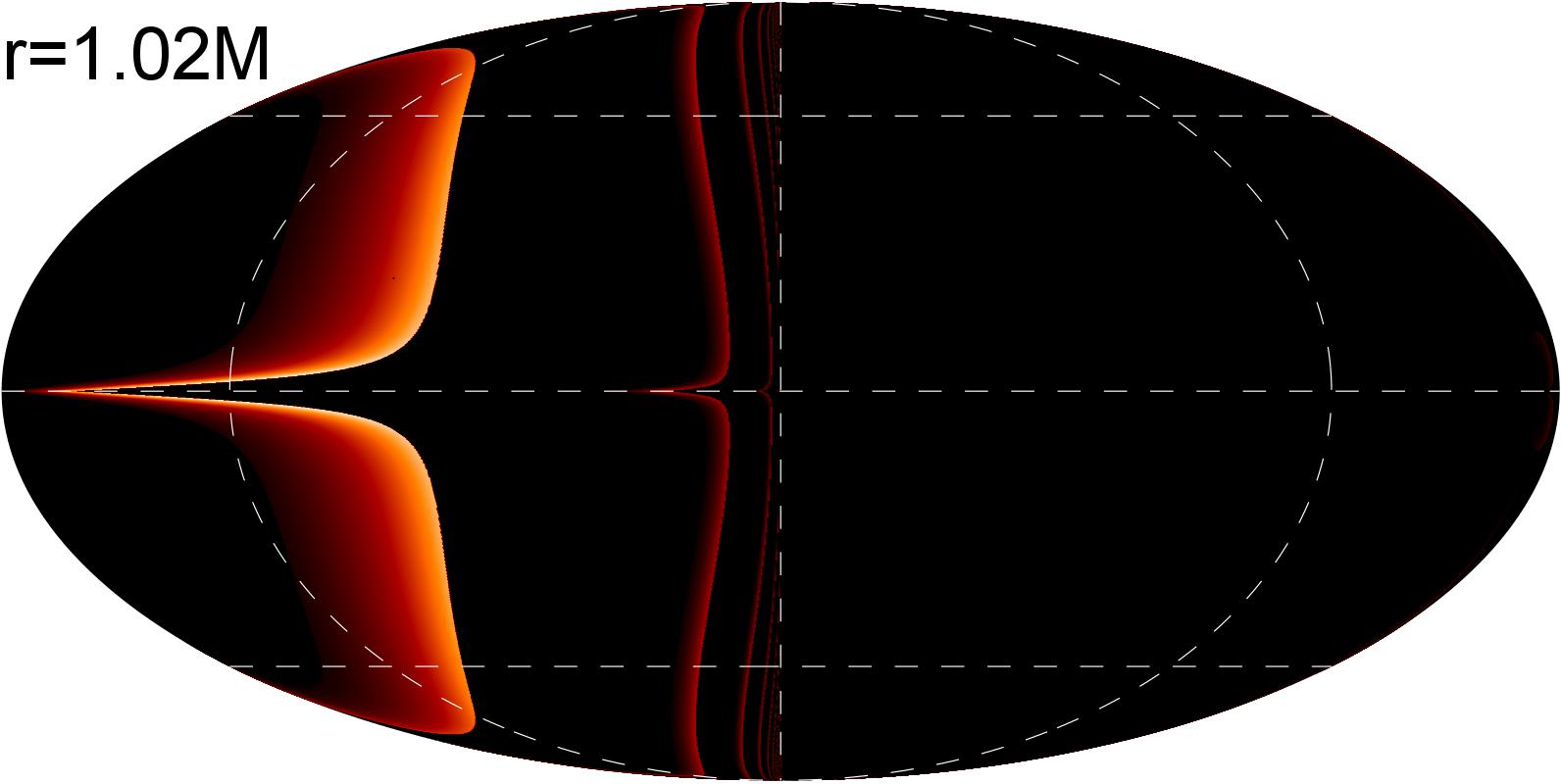}}
\end{center}
\end{figure}

In this case, the incoming flux will not be limited to the local
blackbody radiation from the disk, but will in fact include the entire
view of the disk, as seen by an observer just above or below the
surface. In Figure \ref{fig:diskfig} we show what this disk might
look like from the perspective of an observer at a few different
distances from the black hole. 

The polar projections for these type of plots can be
confusing. Consider the observer on a circular orbit in the black
hole's equatorial plane. The ``forward'' direction defines the
$\phi=0$ azimuth, and the $\theta$ angle is given by the standard
spherical polar coordinates with ``north'' being $\theta=0$ and
``south'' in the $\theta=\pi$ direction. This puts the black hole at
$(\theta,\phi)=(\pi/2,\pi/2)$. Relativistic aberration shifts the
apparent location of the black hole towards $\phi=0$ as the observer
moves to smaller radius. 

Despite the fact that the observer is in the same plane as the
accretion disk, and thus a razor-thin disk would in fact be invisible
in Newtonian physics, the black hole's gravity bends the trajectories
of photons coming from the disk, making it appear as a ring (the
``Einstein ring'') above and below the equatorial plane. In all cases
in Figure \ref{fig:diskfig},
the disk is orbiting in the prograde direction around a maximally
spinning black hole, so the approaching edge on the left side of each
image will appear brighter due to relativistic beaming. Close to the
black hole, multiple images of the disk become visible, due to photons
that orbit the hole many times before eventually escaping and reaching
the observer \citep{James2015a}. 

\begin{figure}[h]
\caption{\label{fig:disk_above} Habitable zone for planets just above
  or below
  a thin accretion disk around a Kerr black hole with $a/M=1$ and mass
  $10^8 M_\odot$ (solid lines, yellow shading) and $10^9 M_\odot$
  (dashed lines, red shading).}
\begin{center}
\scalebox{0.8}{\includegraphics{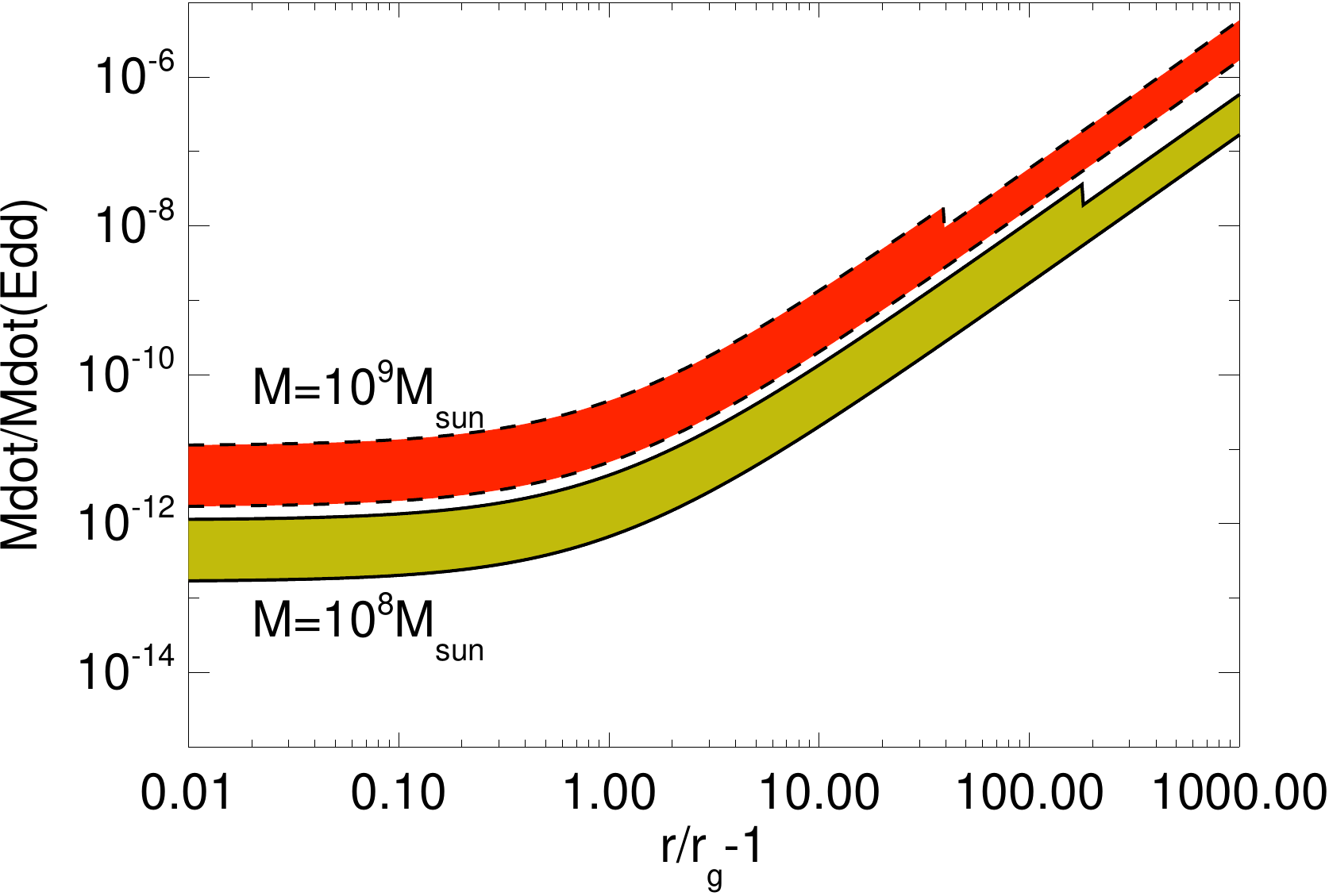}}
\end{center}
\end{figure}

From $r=200M$, the lensed accretion disk is roughly the
same size of the Sun as viewed from the Earth. So if we set the
accretion rate in order to match the apparent blackbody temperature of
$\sim 6000$ K, the planet should be right in the middle of the
HZ. This is exactly what we find, as shown in Figure
\ref{fig:disk_above}, which is identical to Figure
\ref{fig:disk_embed}, only instead of the planet being embedded in the
accretion disk, it is located just above or below the disk. Close to
the black hole, this doesn't make much difference, because the extreme
gravitational lensing makes it look like much of the sky is subtended
by the thermal disk. At larger radii, where the local disk temperature
is much colder, a planet outside of the disk will receive much more
flux from the hot---yet distant---inner regions of the disk. Thus for
a given accretion rate, the HZ moves out to larger radii relative to
the embedded-in-disk paradigm. Also, because the flux is coming mostly
from one specific direction, tidal locking will play an important role
in determining habitability, as discussed above in Section
\ref{section:habitability}. This can be seen clearly in Figure 
\ref{fig:disk_above}, which shows a broader HZ at smaller distances
from the black hole, inside the tidal locking radius. 

Regardless of the specific details---whether the planet is
tidally locked or not, embedded in the disk or not---it is clear that
for any planet to be habitable while also receiving its primary
heating flux from a Novikov-Thorne type accretion disk, the planet
either has to be extremely far from the black hole, or the mass
accretion rate has to be an infinitesimal fraction of Eddington. This
leads to an important inconsistency: The Novikov-Thorne model is
predicated on an optically thick, radiatively efficient disk. At
accretion rates down near $10^{-12} \dot{M}_{\rm Edd}$, the optical
depth is unlikely to be large enough to allow the gas to cool, thus
leading to a very hot, low-density accretion flow.

\subsection{Advective Accretion}\label{section:ADAF}

Radiatively inefficient accretion flows (RIAF) encompass a large
collection of more specific models, generally described by low
densities, high temperatures, and large radial inflow velocities
\citep{Yuan2014}. For simplicity, we adopt a collisionless,
spherically symmetric inflow model as in \citet{Zeldovich1971}. In
this case, the density is given by 
\begin{equation}\label{eqn:n_r}
n(r) = n_0 \left(1+\frac{2GM}{\sigma_0^2 r}\right)^{1/2}\, ,
\end{equation}
where $n_0$ and $\sigma_0$ are the density and velocity dispersion at
large radius, respectively. The bulk velocity of the gas is given by
the geodesic trajectory of a marginally bound particle falling from
infinity with zero angular momentum \citep{Schnittman2015}.

For a planet on a circular planar orbit in the black hole's equatorial
plane, the RIAF acts as a sort of ``head wind'' of low-density gas. We
assume the kinetic energy of this gas is efficiently converted to heat
in the planet's atmosphere, providing the necessary heat for
habitability. The center-of-mass energy of a single RIAF proton with
4-velocity $u_1^\mu$ hitting a proton in the planet's atmosphere with
4-velocity $u_2^\nu$ is 
\begin{equation}\label{eqn:Ecom}
E_{\rm com} = m_p c^2 \sqrt{2(1-g_{\mu \nu} u_1^\mu u_2^\nu)}
\end{equation}
and the kinetic energy available for heat production is simply $E_k =E_{\rm
  com}-2m_p c^2$. Thus the total incoming flux is 
\begin{equation}\label{eqn:F_RIAF}
F = E_k n v\, ,
\end{equation}
with $v$ the velocity of the planet relative to the accretion flow. 

The density at a specific value of $r/r_g$ scales like $M^{-1}$. This
is because the total accretion rate is proportional to density times
area, the accretion rate is proportional to the Eddington-scaled
rate times the black hole mass, and the area is proportional to
$M^2$, so for a fixed Eddington-scaled rate, $\rho(r/r_g) \sim
M^{-1}$. 

\begin{figure}[h]
\caption{\label{fig:HZ_RIAF} Habitable zone for planets surrounded by
  a radiatively inefficient accretion flow around a Kerr black hole
  with $a/M=1$ and mass $10^8 M_\odot$ (solid lines, yellow shading)
  and $10^9 M_\odot$ (dashed lines, red shading).}
\begin{center}
\scalebox{0.8}{\includegraphics{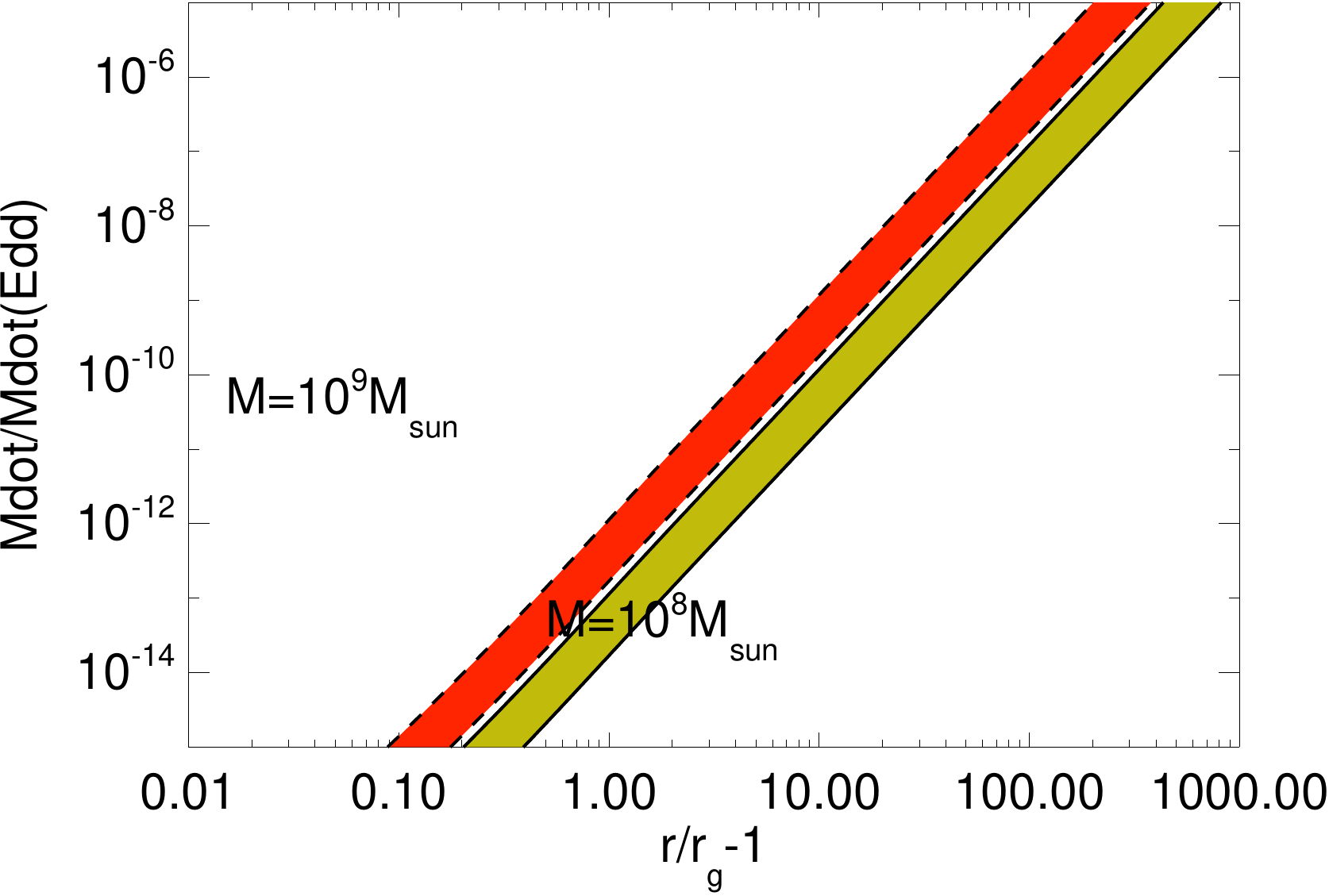}}
\end{center}
\end{figure}

In Figure \ref{fig:HZ_RIAF} we again plot the range of the HZ in terms
of mass accretion rate and orbital radius for a Kerr black hole, now
for a planet getting heated by the kinetic energy of incoming
background gas. At large radius, the RIAF density is low and the
relative velocity between the planet and the inflowing gas is small,
so a higher net accretion rate is required in order to generate
sufficient heat. As one approaches the horizon, the center-of-mass
energy of incoming gas coliding with the planet's atmosphere becomes
so great, anything more than a tiny
fraction of the Eddington accretion rate would be catastrophic for
life on the planet. 

For Miller's planet, orbiting Gargantua at radius
$r/M=4 \times 10^{-5}$, the relativistic boost of incident particles
would be a factor of nearly a thousand. Each gram of gas hitting the
atmosphere would deliver the equivalent energy of 20 megatons of
TNT!\footnote{See {\tt http://what-if.xkcd.com/1/} for a discussion of
  a similar application to the game of baseball.} Furthermore, as we
will see in detail in the next section, the extreme time dilation near
the black hole horizon will greatly increase the apparent rate at
which the accreting gas is hitting the planet, making it even harder
to achieve a habitable environment.

\subsection{Background Radiation}\label{section:background}

Let us imagine, for argument's sake, that there is in fact no
accretion onto the black hole (clearly a deviation from the beautiful
accretion disk images in {\it Interstellar}). In this case, what
energy sources might a habitable planet avail itself of? Fortunately,
everywhere you look in the Universe, there is a constant, steady
stream of background radiation, most notably the cosmic microwave
background (CMB). Shining at a cool 2.7 K, the CMB is hardly a
promising source of radiation to keep a planet in the habitable zone.

\begin{figure}[h]
\caption{\label{fig:time_dilation} Time dilation for an observer at
  the ISCO, as a function of black hole spin ({\it left}) and radius
  $R_{\rm ISCO}$ ({\it right}).}
\begin{center}
\scalebox{0.45}{\includegraphics{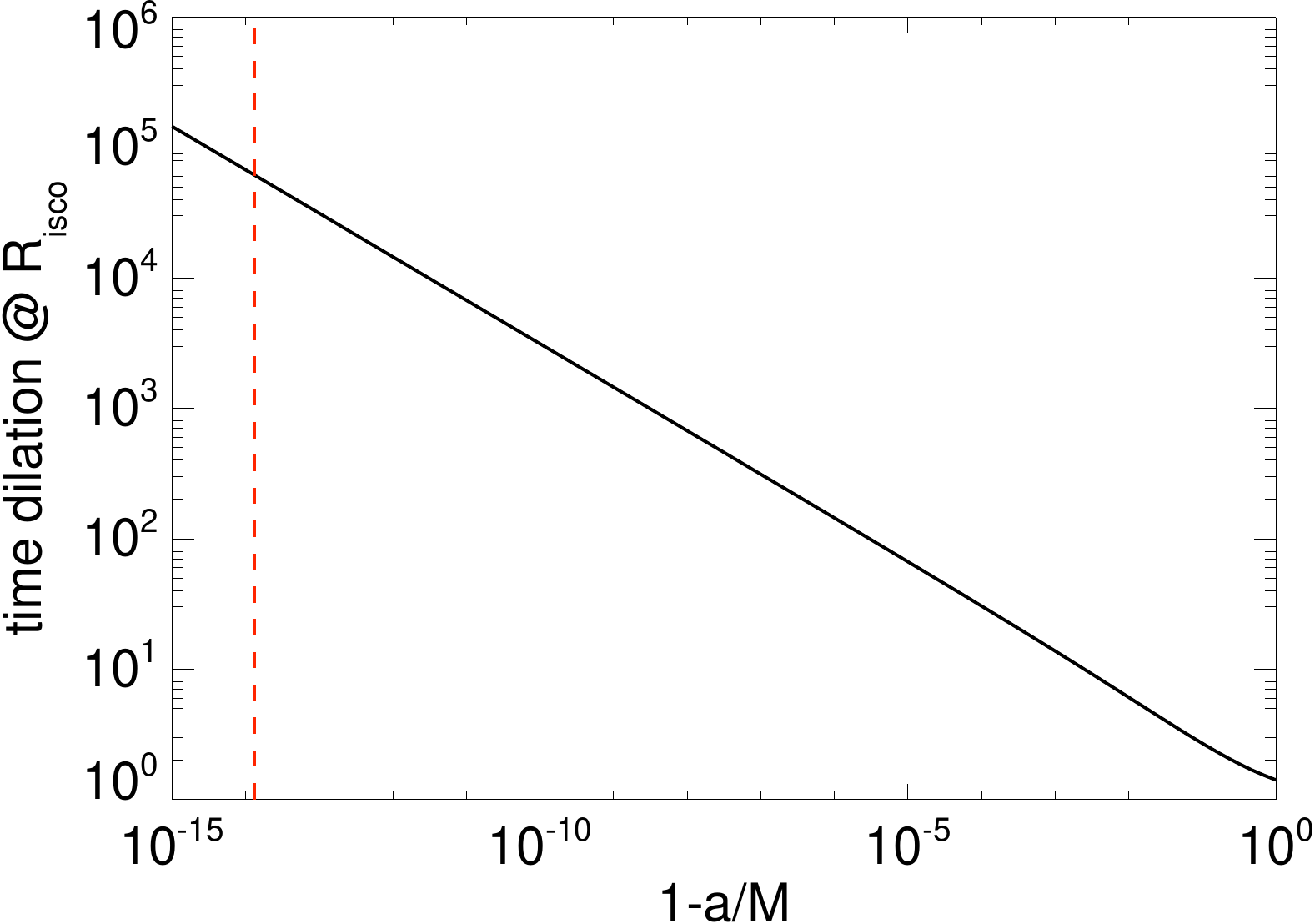}}\
\hspace{0.1cm}
\scalebox{0.45}{\includegraphics{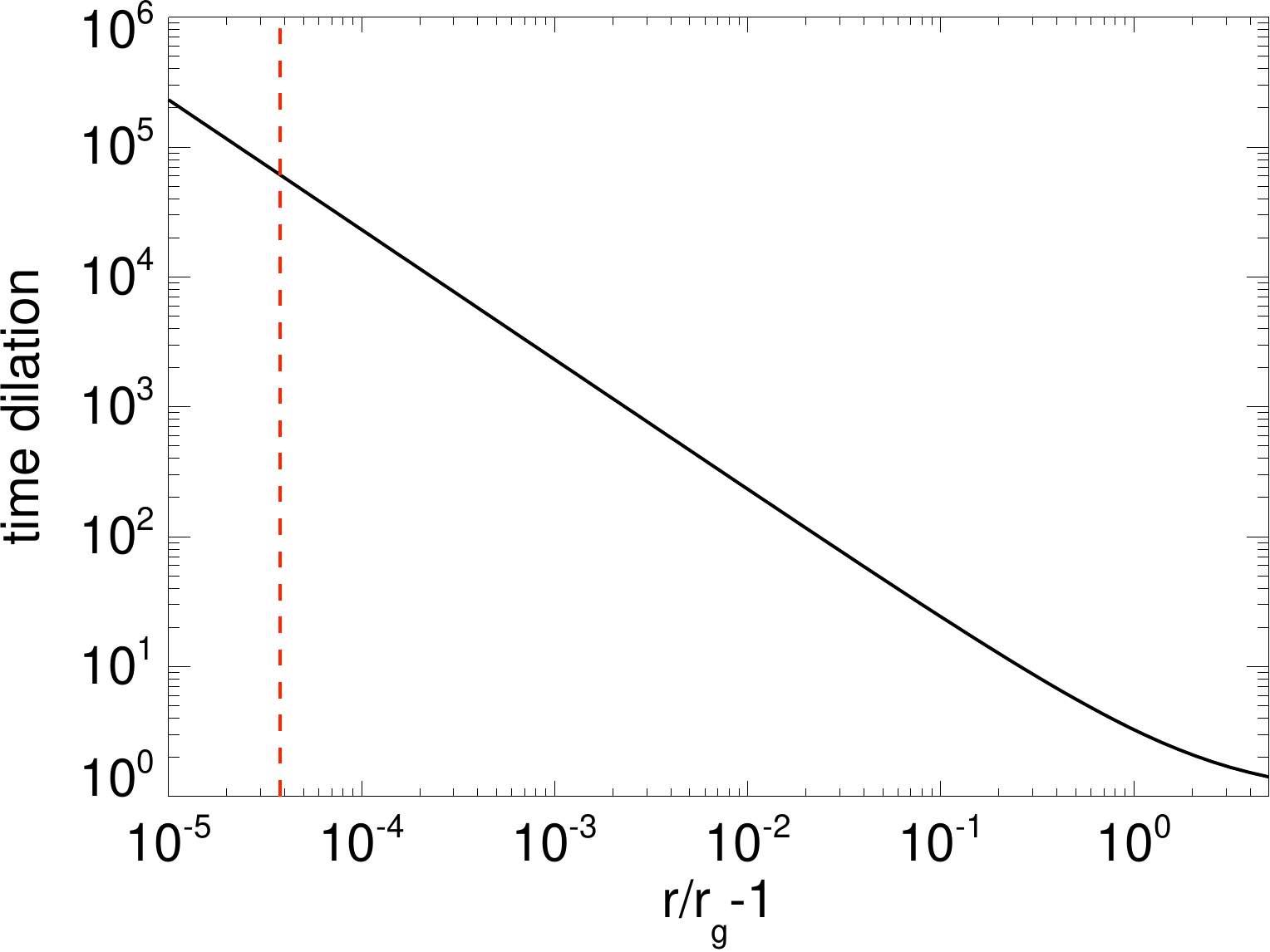}}
\end{center}
\end{figure}

Yet this is where an important feature of relativity comes into
play. As discussed extensively in the film, time slows down for
observers close to a black hole. The time dilation on Miller's planet
is so extreme that one hour corresponds to seven years back on Earth
(or even at the relatively nearby distance of Romily orbiting in the
Endurance mother ship). Assuming the planet is on a stable, circular
equatorial orbit, the minimum distance from the horizon is a strong
function of black hole spin. As shown in Figure
\ref{fig:time_dilation}, Miller's planet---with a time dilation factor
of roughly 60,000---must be at a distance of no
more than $4\times 10^{-5}$ gravitational radii outside the horizon,
and thus the spin must be at least $a/M \gtrsim 1-10^{-14}$
\citep{Thorne:2014}. 

Time dilation is directly related to redshift. Imagine Romily sending
a Morse code message from the Endurance down to the surface of
Miller's planet. If he taps out one beep per second from his point
of view, they will receive 60,000 beeps per second from their point of
view! One can think of a photon as a type of clock, oscillating with
a specific frequency\footnote{Indeed, this is exactly how atomic
  clocks work---connecting our standard of time to that of lasers
  tuned to specific atomic energy transitions.}. As that photon
approaches the black hole, a local observer will measure a higher and
higher frequency. 

\begin{figure}[h]
\caption{\label{fig:cmbfig} All-sky view of the cosmic microwave
  background radiation, as seen from an observer orbiting in the
  equatorial plane of a Kerr black hole with spin $a/M=1$. In each
  frame, the color scale is linear and covers a range from the lowest
  to highest observed CMB temperatures.}
\begin{center}
\scalebox{0.5}{\includegraphics{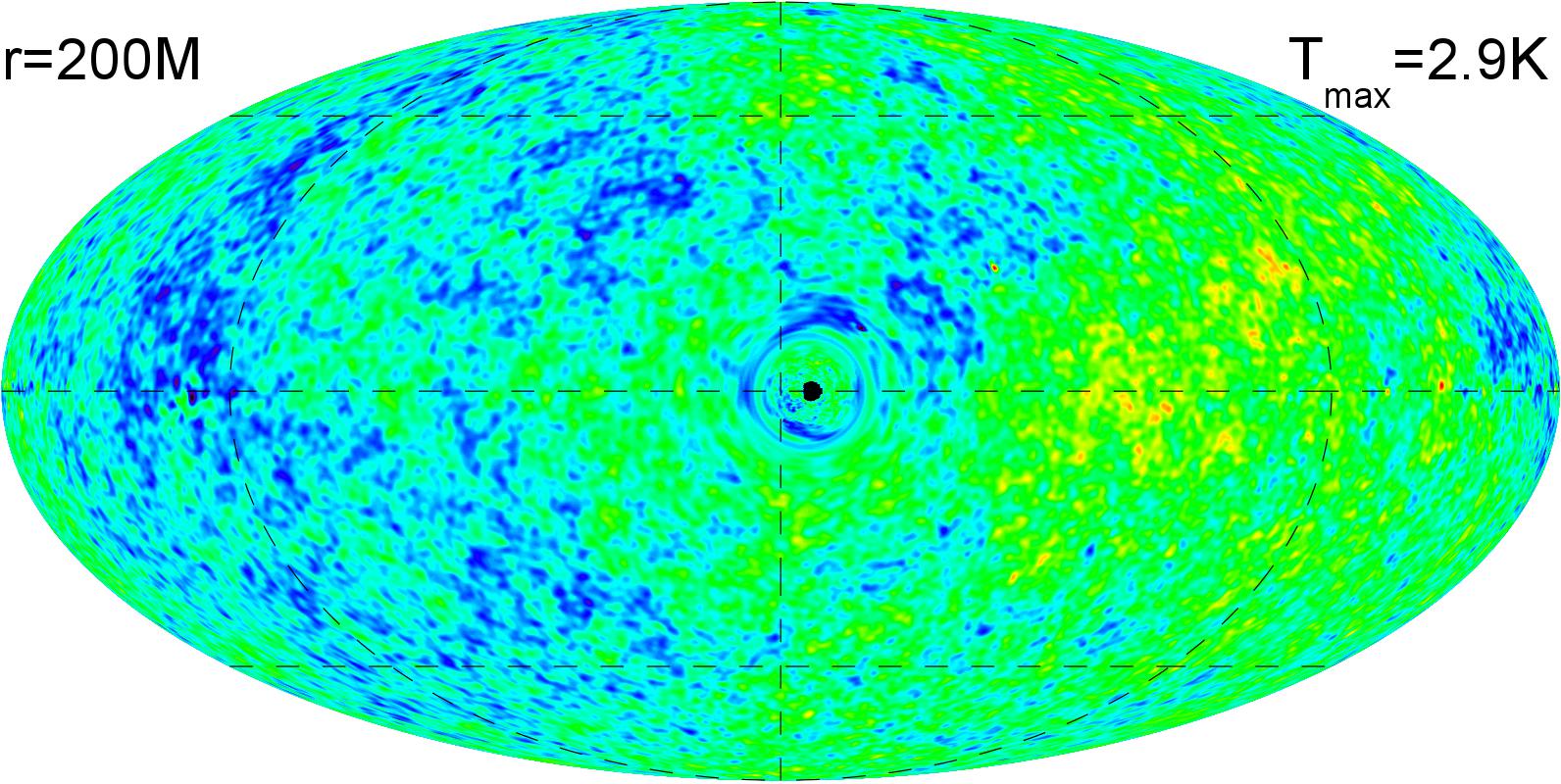}}\
\hspace{0.2cm}
\scalebox{0.5}{\includegraphics{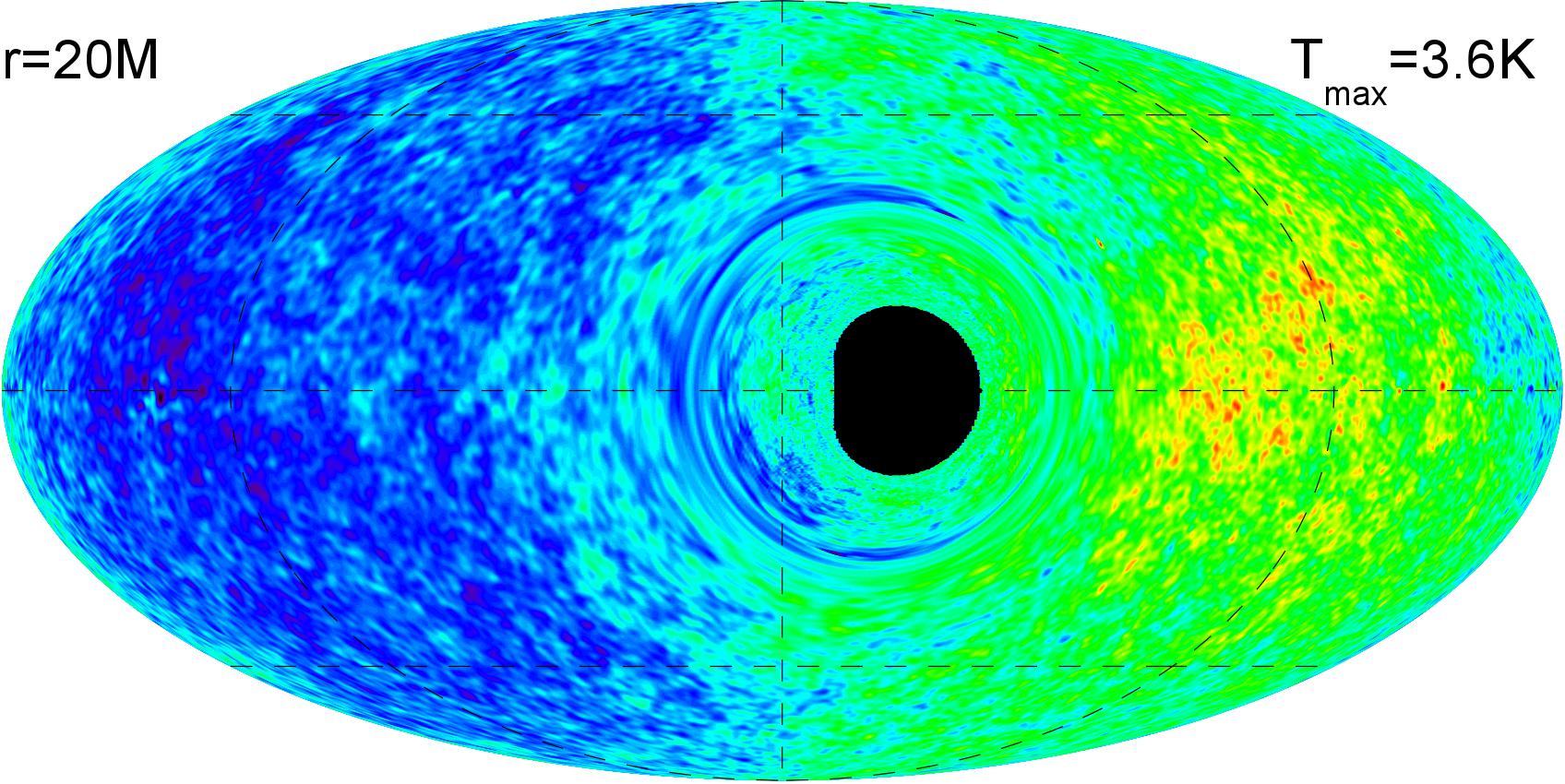}}\\
\vspace{0.2cm}
\scalebox{0.5}{\includegraphics{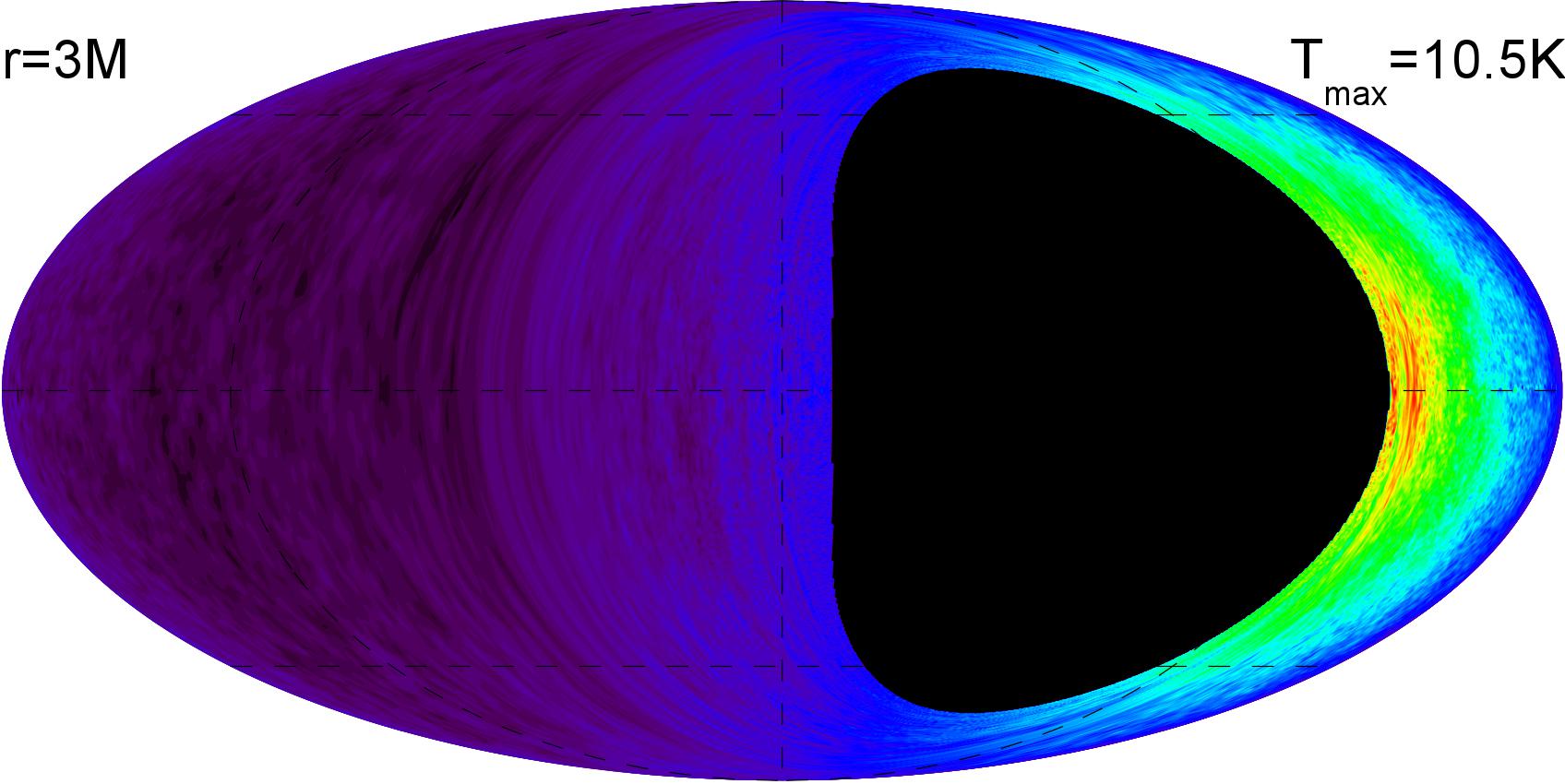}}
\hspace{0.2cm}
\scalebox{0.5}{\includegraphics{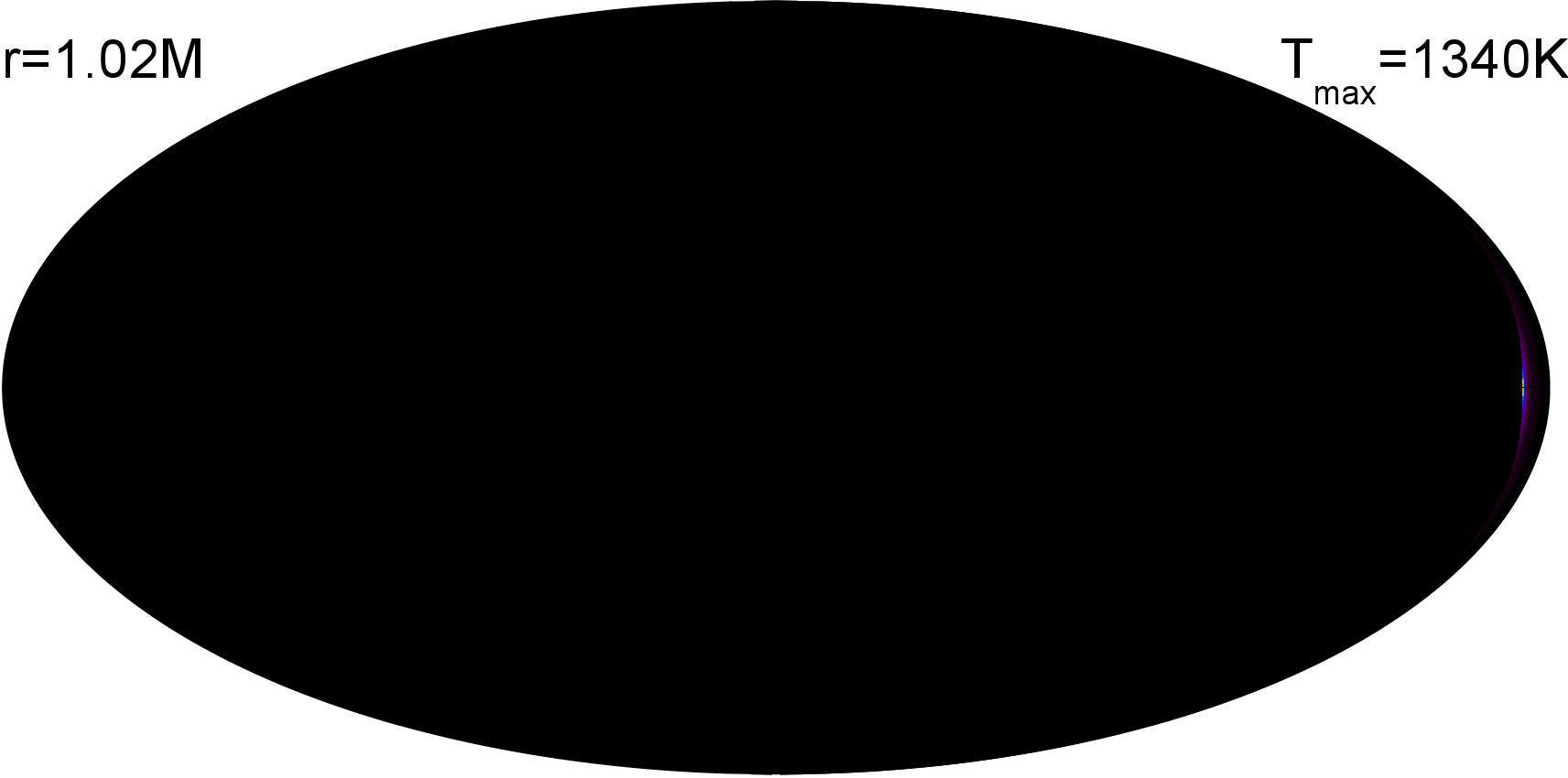}}
\end{center}
\end{figure}

One of the nice things about blackbody spectra is that
temperature is directly proportional to frequency, so if you
Doppler-shift the entire spectrum to higher frequency, it remains a
blackbody, only corresponding to a higher temperature. So for an
observer orbiting a black hole, the blackbody CMB will still appear as
a thermal spectrum, with temperature proportional to the blueshift in
each direction. In Figure \ref{fig:cmbfig} we show what the CMB sky
would look like to an observer orbiting a Kerr black hole at a few
different radii. In each frame, the colors are chosen to cover the
range of temperature from the minimum to maximum with a linear
scale. We also intentionally use the standard WMAP color table, even
though it counter-intuitively uses violet for low temperatures and
red-orange for high temperatures. 

Far from the black hole, the asymmetry is dominated by a simple dipole
moment due to the orbital motion of the observer, just as in CMB sky maps
made from Earth. We can also see the Einstein ring around the black
hole, and the unique shadow of a Kerr black hole: an off-center
truncated circle \citep{Chandra:1983}. Closer to the black hole, the
size of the black hole shadow steadily increases until it fills half
of the sky. Doppler and gravitational redshifts become more extreme,
as does the relativistic beaming, effectively shrinking the size of
the blue-shifted region, already only a few degrees across in the
bottom-right panel of Figure \ref{fig:cmbfig}. 

\begin{figure}[h]
\caption{\label{fig:cmbzoom} Zoom-in view of the region of greatest
  blueshift. In the left panel, the observer is at $r=1.0004M$ and
  the zoom-in region is roughly the angular size of the Sun as seen
  from the Earth. On the right panel, the observer is on the same orbit as
  Miller's planet ($r=1.00004M$), with the CMB blueshifted to over
  700,000K yet with an angular size comparable to Mars.}
\begin{center}
\scalebox{0.5}{\includegraphics{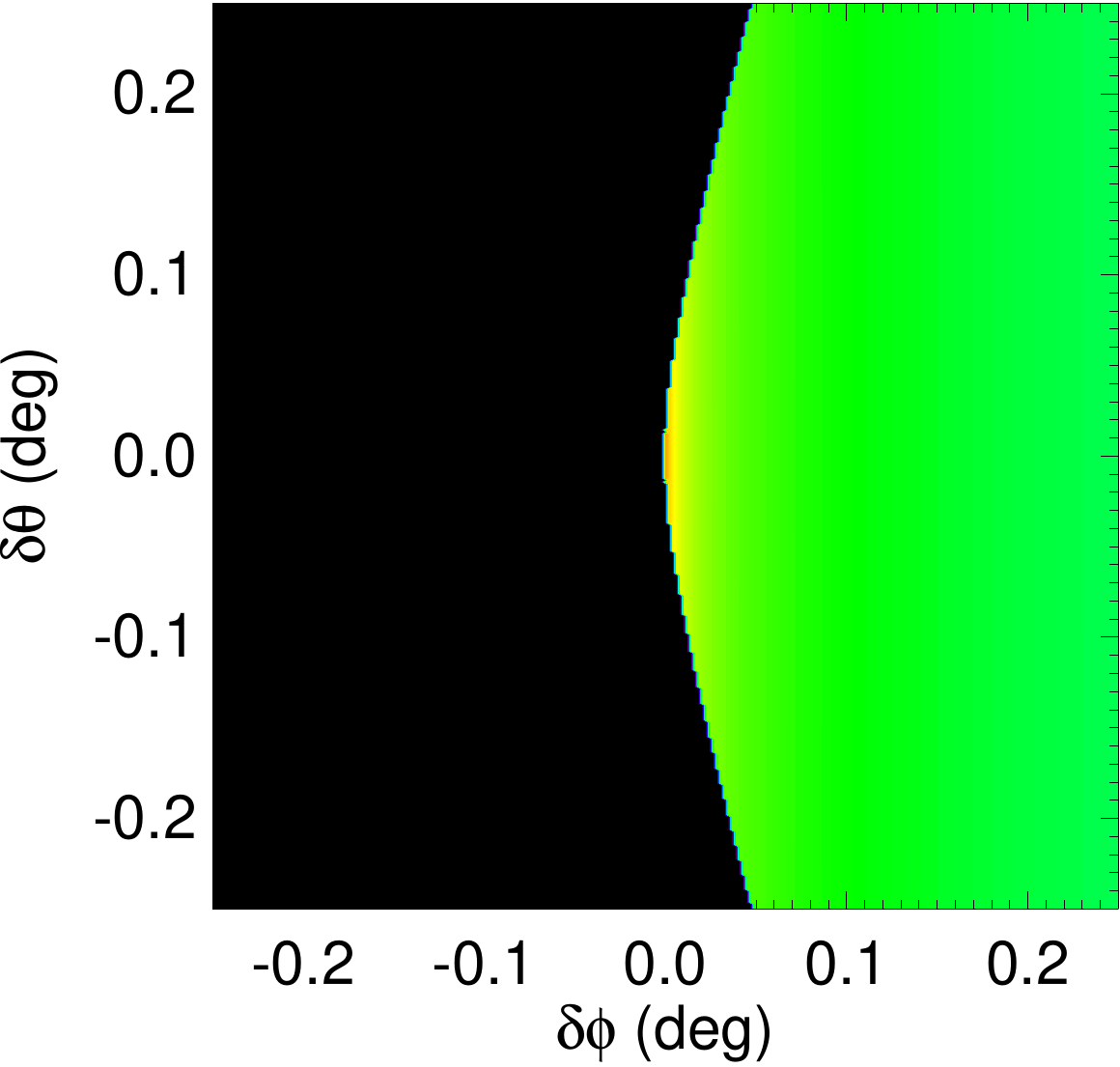}}\
\hspace{0.2cm}
\scalebox{0.5}{\includegraphics{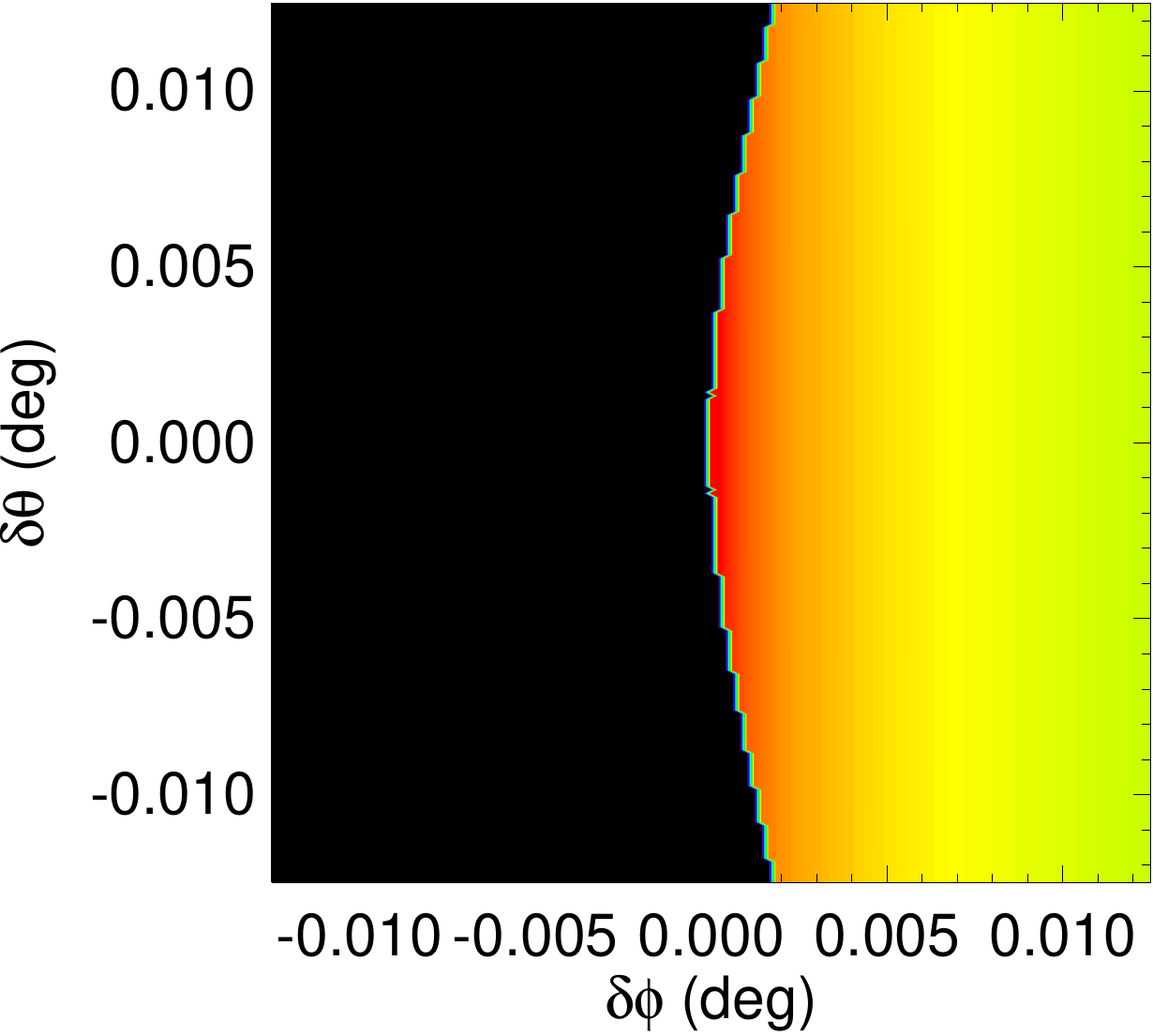}}
\hspace{0.2cm}
\scalebox{0.6}{\includegraphics{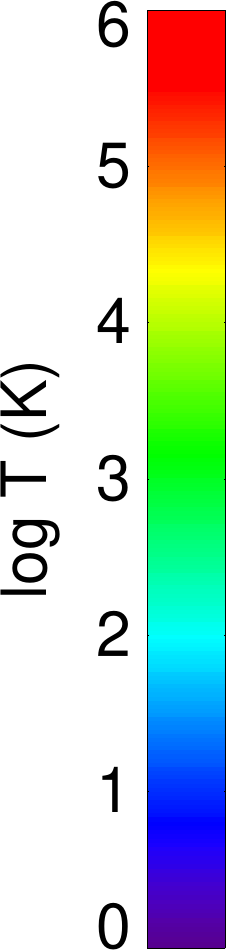}}
\end{center}
\end{figure}

At the same time,
the solid angle of the hot spot decreases like the inverse square of
the time dilation $dt/d\tau$, while the peak temperature approaches
$T_{\rm peak}/T_{\rm CMB} = 9/2 (dt/d\tau)$. These effects are shown in
Figure \ref{fig:cmbzoom}, which zooms in on the area of peak CMB
blueshift at orbital radii of $r=1.0004M$ and $r=1.00004M$ (the
orbital distance of Miller's planet from Gargantua). 

Thus the total incoming flux on the planet scales like $(dt/d\tau)^2
\sim (r-r_{\rm hor})^{-2}$. When the background source is the
isotropic CMB with a temperature of 2.7K, the habitable zone occurs
around $r=1.00036M$, where the incoming CMB has a temperature
of 30,000K. This would be like orbiting a white dwarf at a distance
of 0.2 AU. Perhaps the right temperature for liquid water, but with a
lot of potentially lethal UV radiation.

At this radius, the planet's time dilation relative to infinity would
be a mere factor of 2,000. This would have the added advantage of
giving the atmosphere and surface of the planet roughly five million
years to cool and stabilize while the distant observer on Earth
witnesses a Hubble time of cosmic evolution \citep{Thorne:2014}. On
the other hand, an hour on the surface of this more habitable planet
would only correspond to a few months for Romily in endurance, and
Murph back home on Earth. Dramatic yes, but probably not enough to
drive the plot of the movie.

In any case, this scenario of an idealized Earth-like environment at
$r=1.00036M$ is ignoring (at least) one crucial fact. While the CMB
dominates the total electromagnetic energy density budget of the
universe as a whole, there are localized regions in space where other
backgrounds will certainly dominate. Near the Earth, the total flux in
visible light from nearby stars gives an effective background
temperature of roughly 3K, coincidentally nearly equal to the CMB. Of
course, this is only because we are situated in the middle of a
galactic disk, where the local stellar density far exceeds that of the
Universe on average. 

\begin{figure}[ht]
\caption{\label{fig:EMflux} Radiation flux incident on a planet
  orbiting an extremal Kerr black hole, due to blue-shifted background
  radiation from the CMB (solid curve) or ambient starlight in the
  galactic center (dashed curve). The HZ flux range is marked by the
  horizontal red lines.}
\begin{center}
\scalebox{0.5}{\includegraphics{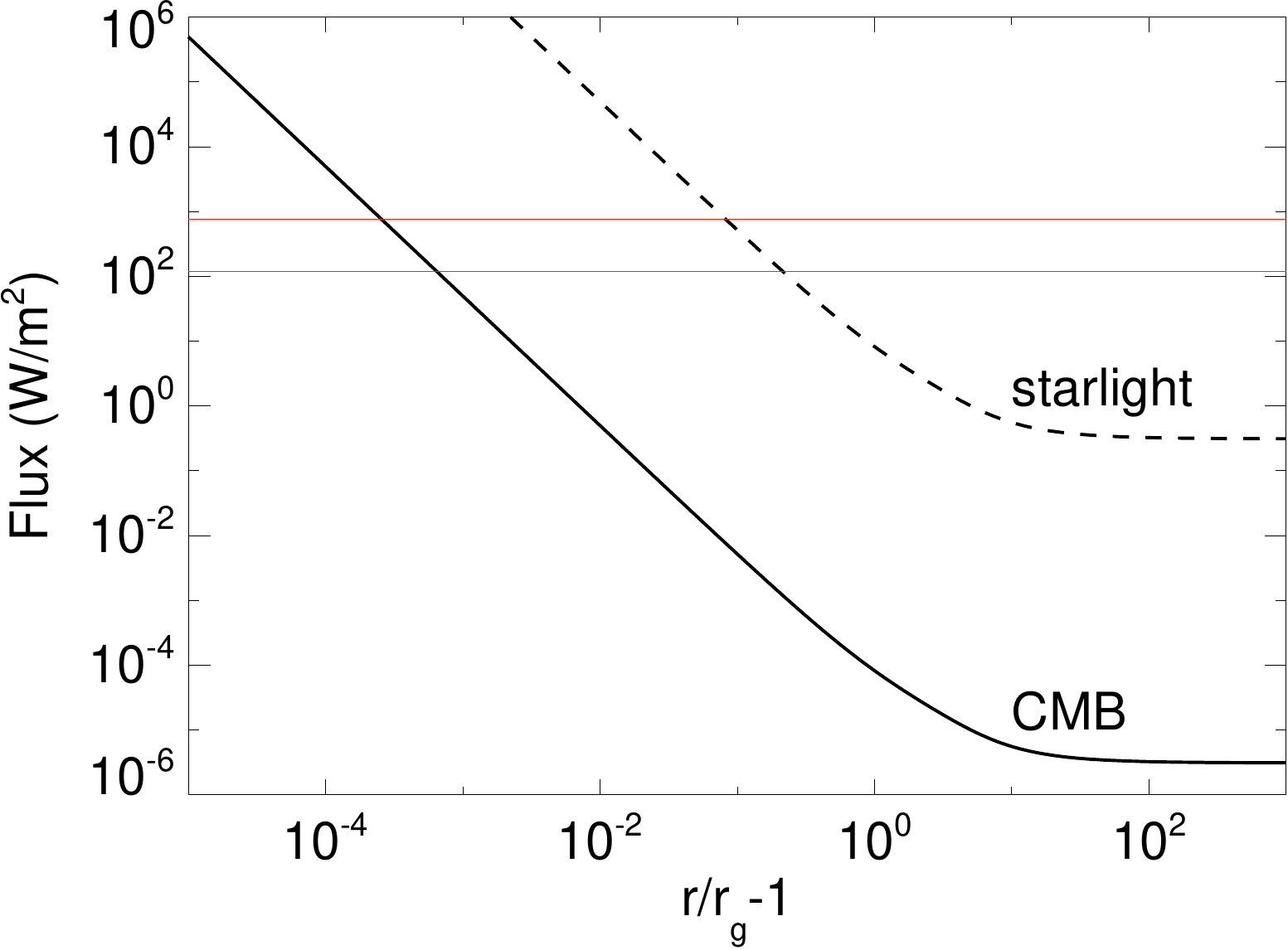}}
\end{center}
\end{figure}

For a planet orbiting a supermassive black hole situated in a galactic
nucleus, the local stellar density will be orders of magnitude
greater. For a planet in our own galactic center, the night sky would
actually be 100,000$\times$ brighter than that of Earth! In Figure
\ref{fig:EMflux}, we plot the mean flux on the planet's surface as a
function of distance from the BH horizon for $a/M=1$. The habitable
zone for a tidally locked planet is marked in red. Note that the HZ
for planets irradiated by blueshifted starlight is at a relatively
large radius of $1.1r_g$, where the time dilation is a measly factor
of 23, but the blackbody temperature of a sun-like star would still be
a whopping 600,000K. Again, technically habitable from an energy
balance point of view, but challenging from a photochemical
perspective. 

\subsection{Neutrinos}\label{section:neutrinos}
Perhaps a civilization that is sufficiently technologically advanced
will be able to construct a sort of ``reverse Dyson sphere''
surrounding their planet with highly reflective material, effectively
raising the planet's albedo to near unity. This would allow
habitability much closer to the host SMBH, even in the face of
overwhelming background UV or X-ray radiation. 

Yet even with such a protective shield, there is still the specter of
nature's silent killer: neutrinos. From a simple calculation of
thermal equilibrium during the early expansion after the big bang, one
can determine that a cosmic neutrino background freezes out at
temperatures of roughly 2.5 MeV. As relativistic particles, the
neutrinos cooled exactly like the photons during the expansion, but
the photons received the added energy stored in the electron-positron
plasma that annihilated around temperature 0.5 MeV. Thus the
subsequent radiation background is slightly hotter than the primordial
neutrino background. While the CMB has a blackbody temperature of 2.73
K, the C$\nu$B has $T=1.95$ K \citep{Weinberg2008}. While this
background has not yet been detected directly, it has been indirectly
inferred from the CMB measurements with Planck \citep{Planck2015}.

Of course, the much more important difference between the neutrino and
photon backgrounds is that the neutrinos have a very small cross
section, so most go right through the atmosphere without depositing
any energy. A reasonably good estimate for the neutrino-baryon cross
section can be approximated by
\begin{equation}
\sigma_\nu \approx 9 \times 10^{-44} \mbox{cm}^2 \left(\frac{E}{\rm{MeV}}\right)^2
\end{equation}
for neutrino energies below 100 MeV, and 
\begin{equation}
\sigma_\nu \approx 7 \times 10^{-42} \mbox{cm}^2
\left(\frac{E}{\rm{MeV}}\right)
\end{equation}
at higher energies. 

\begin{figure}[ht]
\caption{\label{fig:cnub} Neutrino flux absorbed by the atmosphere
  (solid line) or body (dashed line) of an Earth-like planet
  orbiting an extremal Kerr black hole, due to blue-shifted cosmic
  neutrino background (C$\nu$B).}
\begin{center}
\scalebox{0.5}{\includegraphics{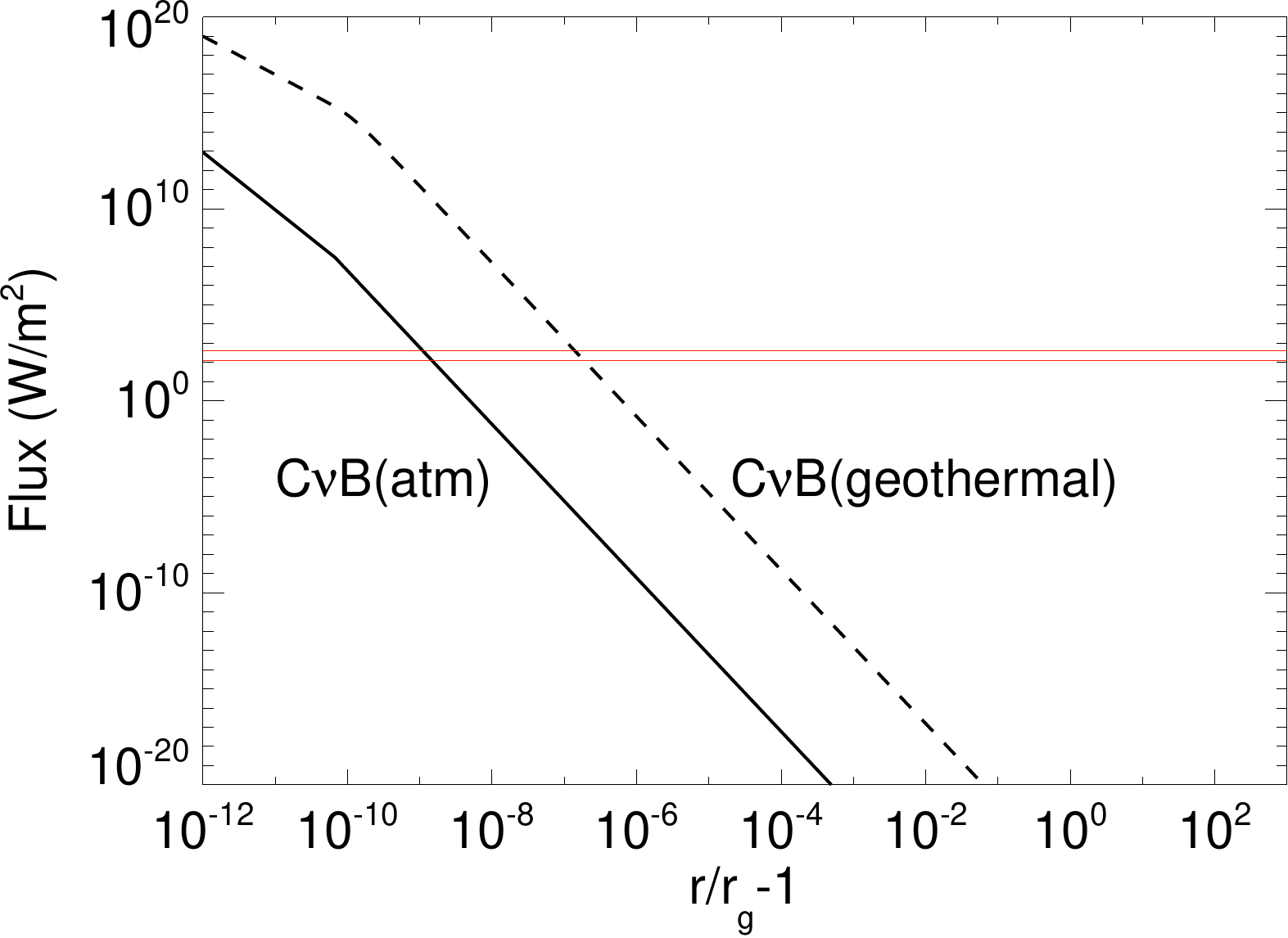}}
\end{center}
\end{figure}

The blue-shifted C$\nu$B flux is shown in Figure \ref{fig:cnub} as a
function of radius. Note the vastly expanded axes relative to Figure
\ref{fig:EMflux}. Only at a blueshift of roughly 10 billion does the
C$\nu$B deposit enough energy in the atmosphere to keep a planet
within the habitable zone (designated by the horizontal red lines, as
in Fig.\ \ref{fig:EMflux}). Yet many of the neutrinos that pass right
through the atmosphere are subsequently absorbed in the dense core of
the planet, so this energy could also be harnessed for heating up the
surface for life. In that case, the required blueshift for
habitability is 100 times smaller (dashed line in
Fig.\ \ref{fig:cnub}). Note the bends in the flux curves around
$r/r_g=1+10^{-10}$ correspond to the point where the typical C$\nu$B
energy rises above 100 MeV, where the cross section scaling
changes. It coincidentally also corresponds to the point where the
planet's body becomes optically thick to neutrinos, and thus the
absorbed flux doesn't rise as much with increasing energy and cross
section, giving a slightly flatter curve than that of the flux
absorbed in the atmosphere.

On Earth, the C$\nu$B flux (incident, not absorbed) is slightly
smaller than that of the CMB, with a total flux of $8\times 10^{-7}$
W/m$^2$. On the other hand, the solar neutrino flux is a whopping 40
W/m$^2$, and these neutrinos have much higher energies---typically
$0.1-10$ MeV, giving much higher cross sections. Of course, a planet
like Miller's planet won't have a nearby star bathing it in neutrinos
from nuclear reactions. In the interstellar medium, the dominant
source of high-energy neutrinos is the background population of old
supernova remnants (SNRs), contributing roughly $10^{-8}$ W/m$^2$ of
neutrinos with energies of 10 MeV and above \citep{Sigl:2012}. We
refer to this source as the Galactic neutrino background (G$\nu$B).

\begin{figure}[ht]
\caption{\label{fig:gnub} Neutrino flux absorbed by the atmosphere
  (solid line) or body (dashed line) of an Earth-like planet
  orbiting an extremal Kerr black hole, due to galactic background
  neutrinos. In panel (a) we show flux
  corresponding to Solar System environments, while (b) is appropriate
  for a galactic center with flux roughly $10^5$ times higher.}
\begin{center}
\scalebox{0.45}{\includegraphics{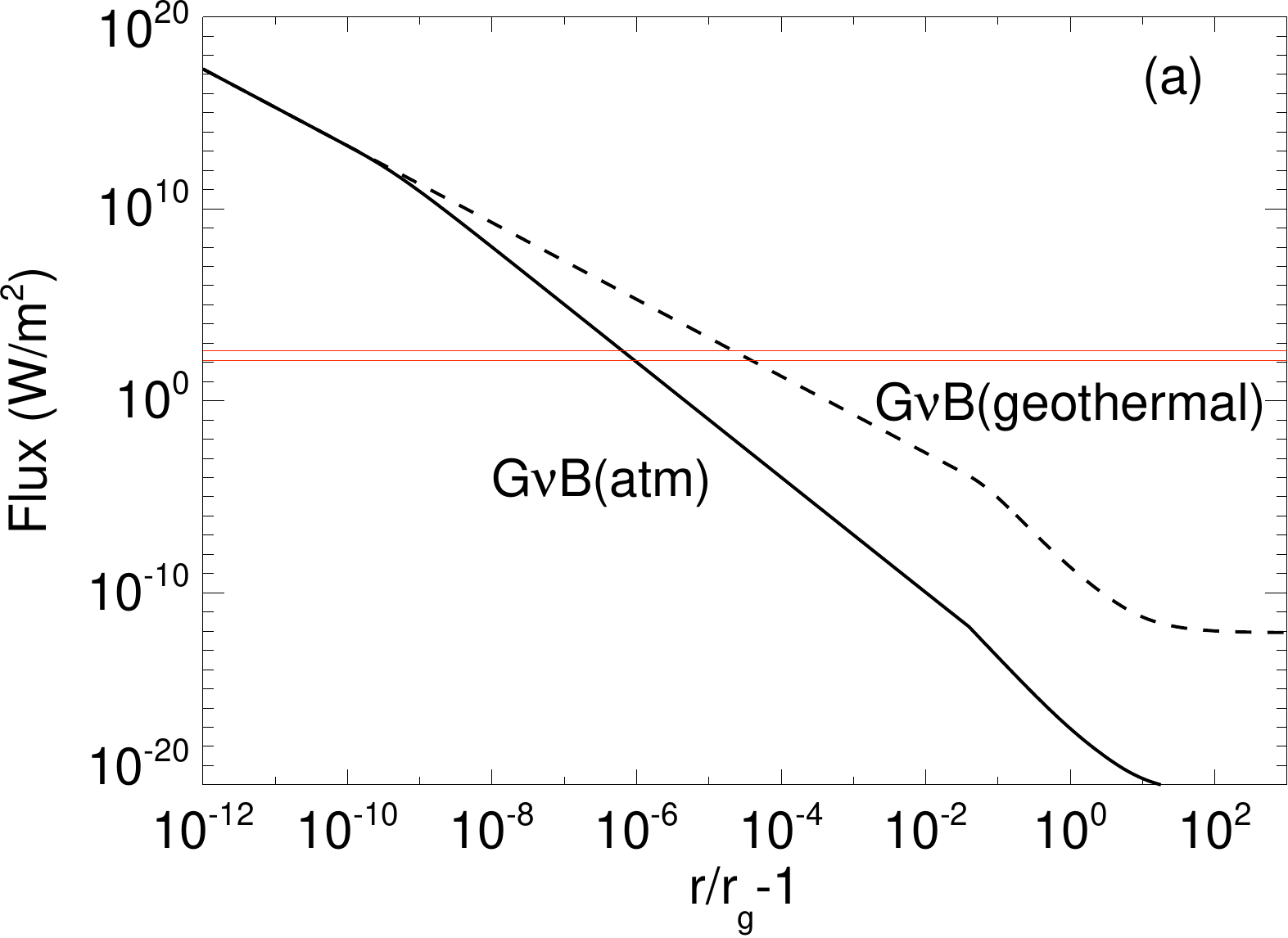}}
\scalebox{0.45}{\includegraphics{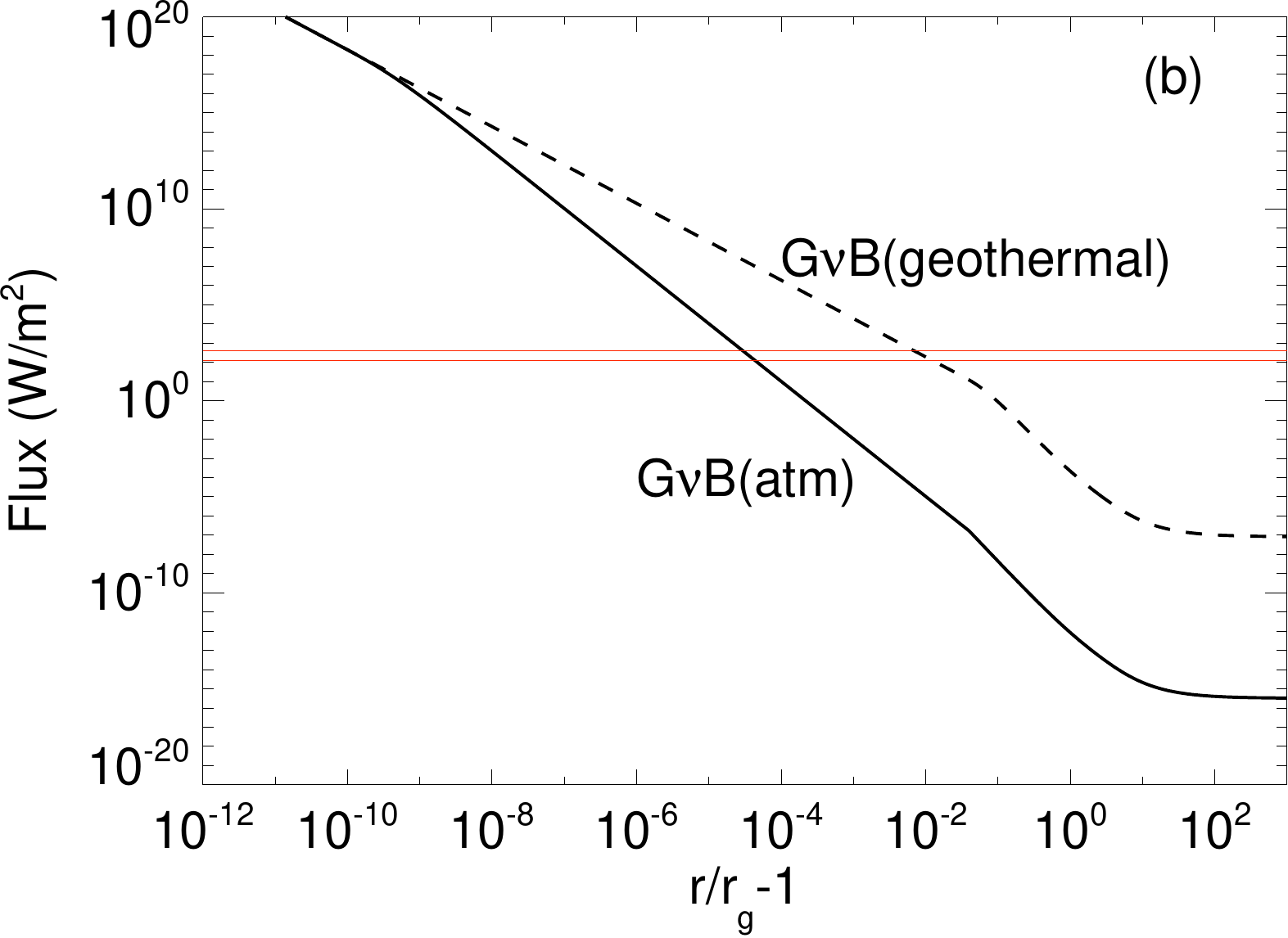}}
\end{center}
\end{figure}

As shown in Figure \ref{fig:gnub}, when blueshifted to
higher energies and cross-sections, this G$\nu$R neutrino flux reaches
habitability levels at $r/r_g-1=10^{-6}$ and $3\times 10^{-4}$ for
atmospheric and planetary absorption, respectively. For these
neutrinos, even the atmosphere is optically thick at
sufficiently high blueshift, corresponding to PeV energies. 
Lastly, we note
that, just like the background starlight,
the SNR neutrino background is expected to be roughly
$10^5$ times greater near the galactic center. These enhanced fluxes
are shown in panel (b) of Figure \ref{fig:gnub}. Remarkably, the
subsequent geothermal heating from neutrinos is only a factor
of 100 below that of the background starlight. And unlike the harmful UV or
X-ray flux from this blueshifted electromagnetic radiation, neutrino
heating of the planet's core could lead to a thriving population of
lifeforms similar to those found near deep ocean vents on Earth.

\subsection{Dark Matter}\label{section:darkmatter}

Neutrinos aren't the only pervasive, invisible source of energy in the
Universe. There is also dark matter, which actually makes up
significantly more of the universe by mass (27\% for dark matter, vs
0.003\% for neutrinos), but is even more difficult to detect. One of
the leading models for dark matter is that of WIMPs: weakly
interacting massive particles. Here ``weakly'' is a technical term,
not just an adjective. It means the invisible particles interact with
regular matter via the weak nuclear force, similar to neutrinos. And
by ``massive,'' we really only mean that the rest mass is much greater
than the energy, as opposed to neutrinos, which are thought to have a
tiny rest mass, and thus their energy is dominated by their
relativistic velocity. 

Most WIMP searches have focused on the GeV-TeV mass scales, so roughly
on the order of typical atomic nuclei. This means that the number
density of WIMPs in the solar neighborhood is actually much smaller
than that of neutrinos, roughly $10^{-3}$ cm$^{-3}$ vs $\sim 100$
cm$^{-3}$ for the neutrinos. Furthermore, because the dark matter (DM)
is cold, the kinetic energy density due to the random motion of the DM
particles (typically $\sim 100$ km s$^{-1}$ in the solar neighborhood)
is also small, on the order of a few eV cm$^{-3}$. In order to be a
viable heating source, we need a much higher concentration of much
higher energy DM particles.

Fortunately, rapidly spinning SMBHs can provide both! After all, just
about the only thing we know for sure about dark matter is that it
interacts with gravity. What better tool could be imagined for
accelerating dark matter than a gravity engine like a black hole?
As we've seen
multiple times in the previous sections, the extreme blueshifts
experienced close to the event horizon can transform insignificant
background radiation into a powerful energy source for sustaining (or
destroying) life on a planet very close to the black hole. The same
goes for dark matter, although now there is an added effect due to the
gravitational focusing of the DM particles by the BH. As described in
\citet{Schnittman2015}, we can treat the DM as made up of two
distinct populations: the bound, and unbound particles. The unbound
particles essentially plunge in towards the BH from far away, and have
total energy close to their rest mass energy. Here, ``far away'' means
just outside the BH influence radius, where the velocity dispersion of
the ambient stars, gas, and DM is dominated by the overall
graviational potential of the galactic nucleus, not the black hole. In
practice, this is typically on the order of parsecs for SMBHs with
masses of $\sim 10^8 M_\odot$, or 200,000 $r_g$.

For the unbound population, the gravitational focusing is a relatively
weak effect, leading to a density profile scaling like $\rho \sim
r^{-1/2}$ \citep{Schnittman2015}. On the other hand, over hundreds of
millions of years, as the SMBH grows through accretion, orbiting DM
particles can get adiabatically trapped on closer and closer orbits,
leading to a steep density cusp around the black hole
\citep{Gondolo1999,Sadeghian2013,Ferrer2017}, with the density scaling
like $\rho \sim r^{-2}$. This bound population has negative binding
energy, and is also limited to stable geodesic orbits, so cannot
sample the full range of phase space as the unbound population. 

\begin{figure}[ht]
\caption{\label{fig:dmdist} Dark matter distribution around a Kerr
  black hole. The DM density, as measured by an observer on a circular,
  equatorial orbit, is plotted in panel (a). In (b) we show the
  average value of $\beta\gamma$ for the DM particles incident on the
  observer. In both plots we show the bound and unbound DM
  populations, as described in the text.} 
\begin{center}
\scalebox{0.45}{\includegraphics{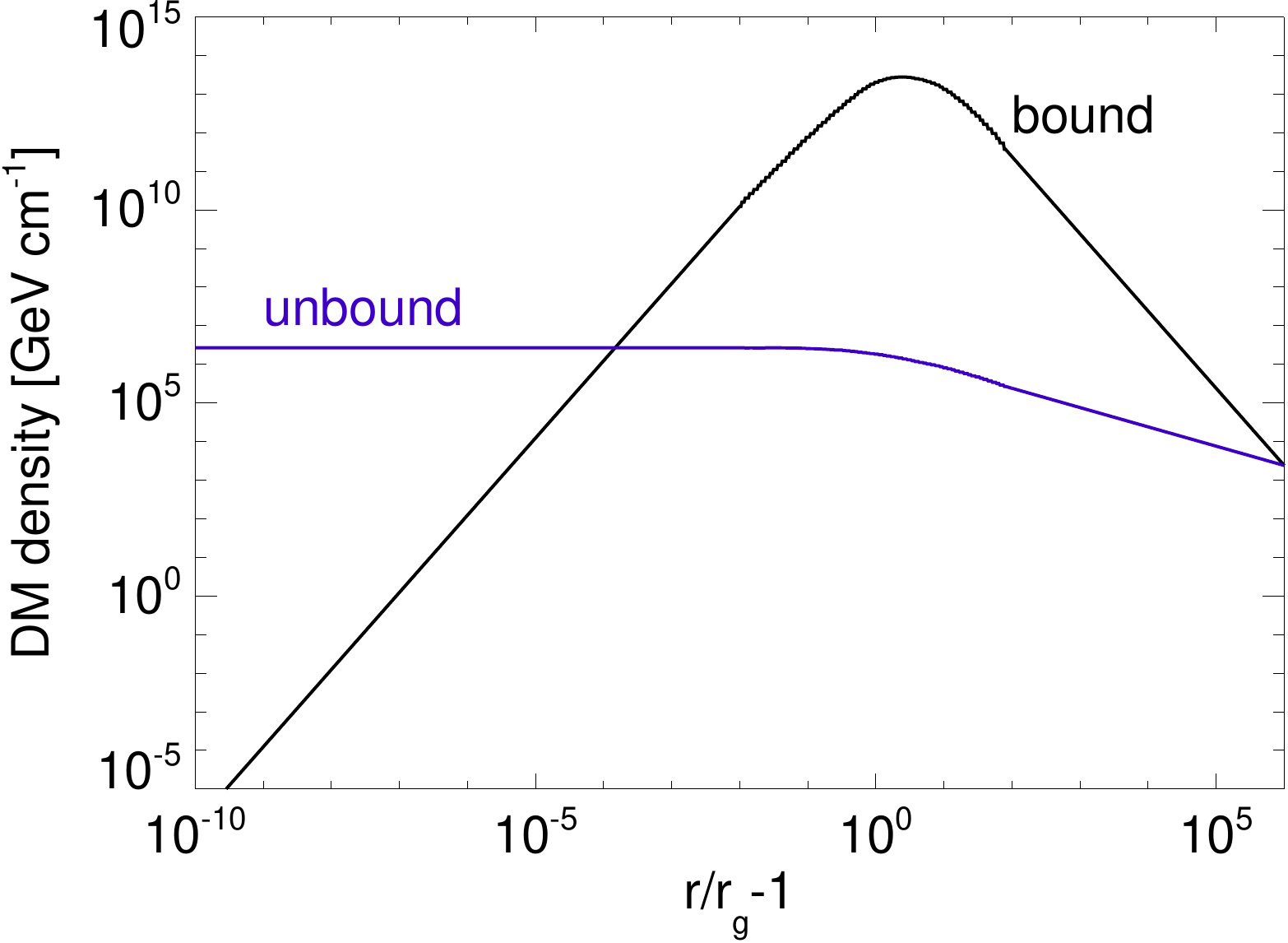}}
\scalebox{0.45}{\includegraphics{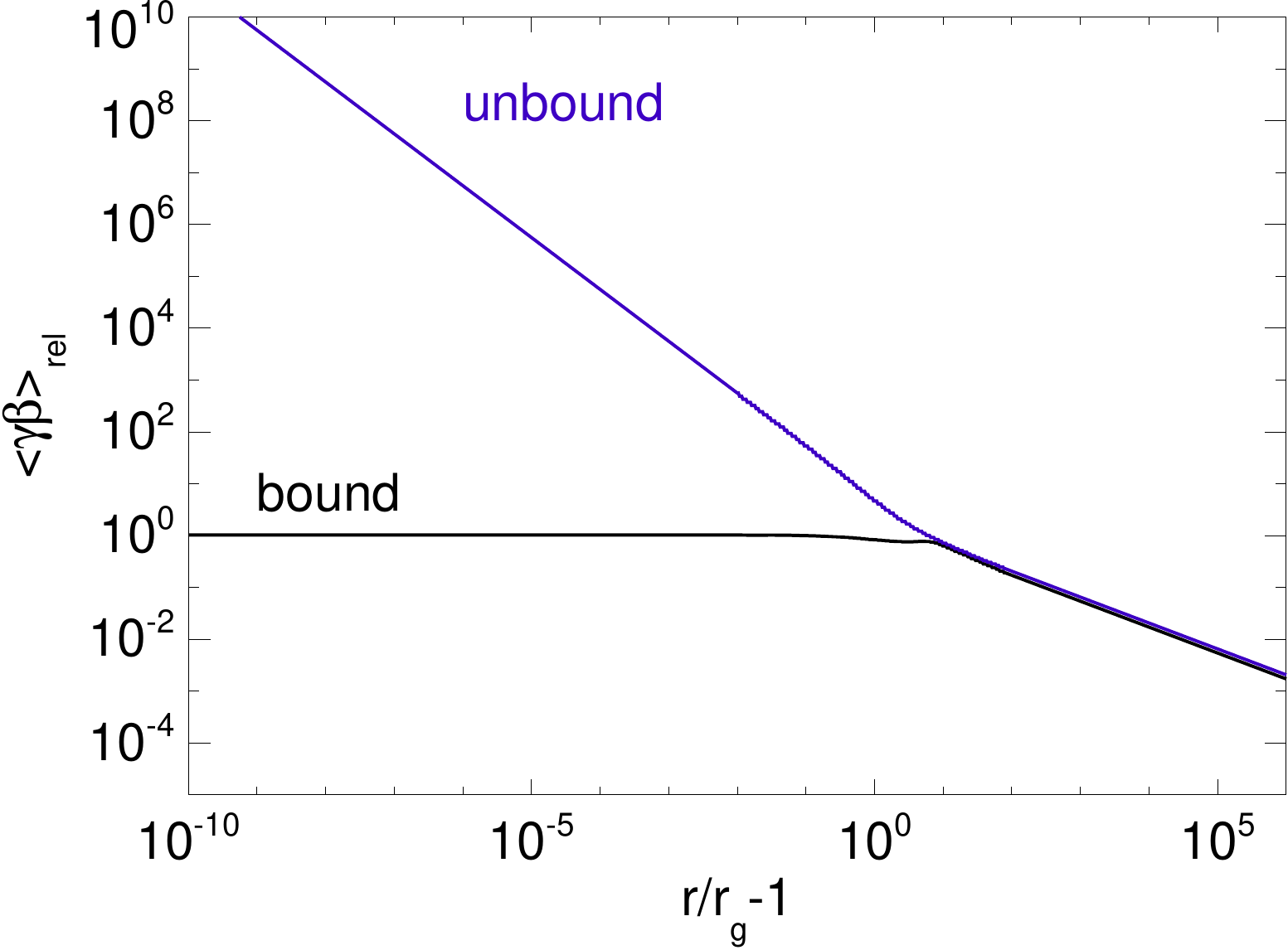}}
\end{center}
\end{figure}

Combining the two effects (gravitational focusing, and random orbital
motion), in Figure \ref{fig:dmdist} we show the
density and mean relative velocity between DM particles as a function
of distance from an extremal Kerr black hole. At large radius, the
populations blend into a single population, but closer to the black
hole we see the bound population's density rise sharply to a peak a few $r_g$
from the horizon, then decrease sharply as fewer and fewer stable
orbits are available. These stable orbits are very nearly circular,
pro-grade, equatorial orbits, so the relative velocity is
constant. For the unbound particles, however, almost any random
trajectory can be reached, leading to extremely high relative
velocities. In fact, for perfectly extremal black holes, the
center-of-mass energy between plunging particles can grow without
bound as the trajectories approach the horizon
\citep{Banados2009,Schnittman2014}. 

\begin{figure}[ht]
\caption{\label{fig:dmflux} Flux on a planet due to scattering of dark
  matter particles off nucleons in the atmosphere (solid lines) or the
  planet's interior (dashed lines). As in Figure \ref{fig:dmdist}, we
  consider both the bound (black lines) and unbound (blue lines) dark
  matter population.}
\begin{center}
\scalebox{0.55}{\includegraphics{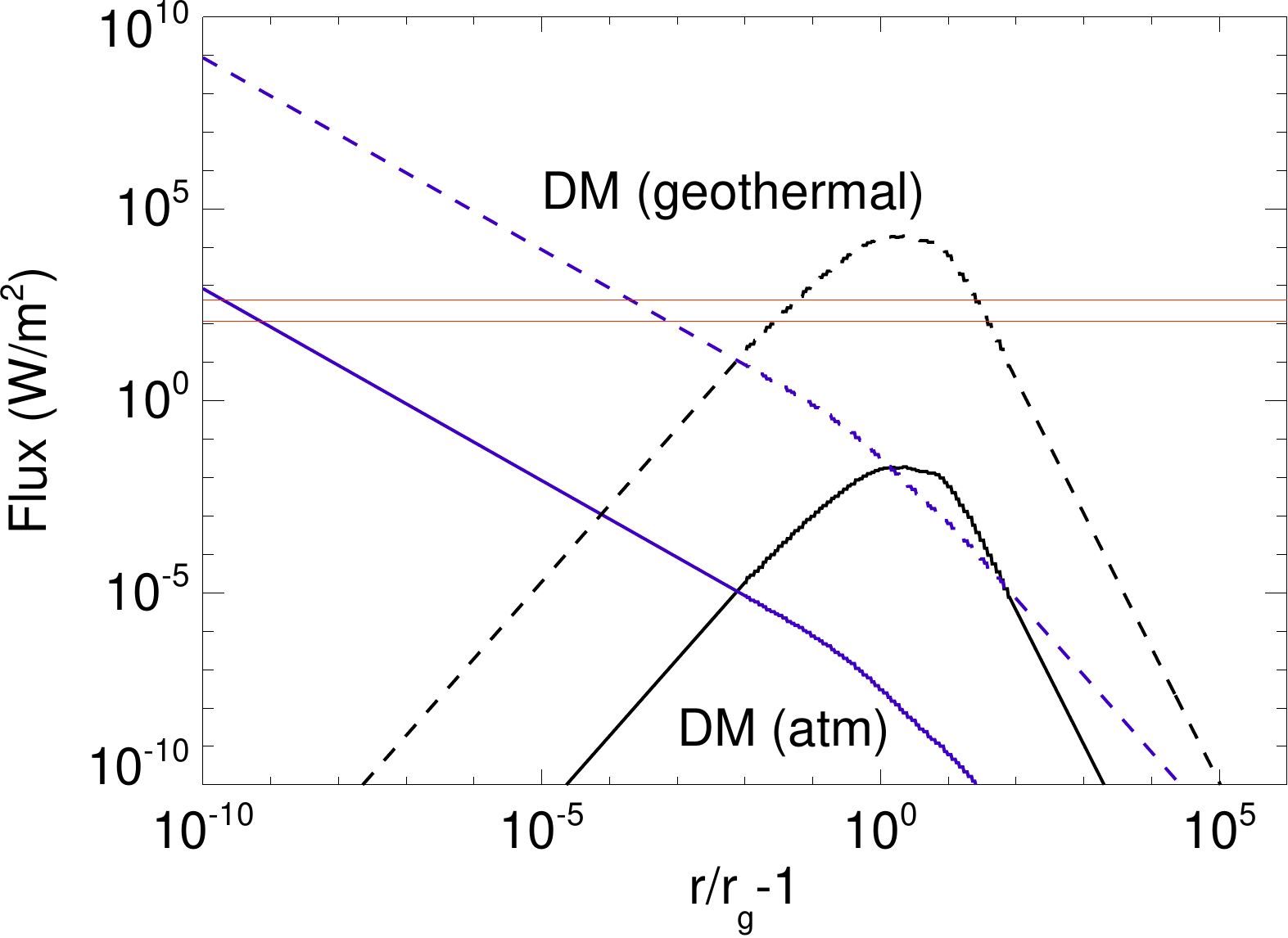}}
\end{center}
\end{figure}

To calculate the heating rate due to DM interactions with a planet, we
need to estimate the cross section of DM-baryon scattering. Since dark
matter has still not been directly detected, we are forced to use the
latest upper limits from underground experiments. We adopt a generous
cross section of $\sigma_{\rm WIMP-baryon}\approx 10^{-41}$ cm$^{-2}$
for $m_\chi = 10$ GeV, based on the results from the CDMS run in
2015 \citep{CDMS2016}, and consistent with other similar underground
experiments. Of course, we really have no idea what the actually cross
section might be, or how it might scale with energy, or how the
energy coupling mechanism might work, but as with the rest of this
paper, we take our best guess, and follow the equations where they
lead (and beyond). 

In Figure \ref{fig:dmflux} we show the energy flux incident on an
Earth-like planet due to WIMP-baryon scattering in the atmosphere
(solid lines) and in the interior of the planet (dashed lines). Ouside
of $r=1.01r_g$, the flux is dominated by the bound population, where
the enhanced density compensates for the lower energy particles. In
fact, the density would be so great as to make the planet uninhabitable
except for two narrow bands at $r\approx 1.03r_g$ and $r\approx 30
r_g$. This is a remarkably distant habitable zone for such a weak (yet
admittedly speculative) energy source. 

We should point out that the very existence of a bound population of
DM and its accompanying density cusp is as-yet unproven. On much
firmer ground is the unbound population, which certainly exists down
to near the influence radius. And as long as there is dark matter near
the influence radius, the gravity of the black hole will certainly
focus and accelerate it as in Figure \ref{fig:dmdist}. In this case, 
the unbound population leads to very high energy collisions inside
of $1.01r_g$, and habitable energy levels around $1.0003r_g$, still
well outside of the nominal orbit for Miller's planet. 

\subsection{Gravitational Waves}\label{section:GW}

According to Kip Thorne, an early version of the screenplay for {\it
  Interstellar} involved the discovery of the wormhole to Gargantua
through the detection of anomalously strong gravitational waves
\citep{Thorne:2014}. Thus we find it fitting to conclude this
exploration by imagining the interaction of gravitational waves (GWs)
with a planet around a SMBH, combining the two most amazing pieces of
Einstein's ``outrageous legacy'' \citep{Thorne:1994}. Ironically,
Thorne's version of the story has LIGO detecting the GWs from a
neutron-star/black hole merger in 2019, which may very well be the
year that the first such event was detected. In ``real life,'' LIGO's
first detection came just over a year after the release of {\it
  Interstellar}, in September 2015.

\begin{figure}[h]
\caption{\label{fig:GWflux} Flux on a planet due to gravitational
  waves from a stochastic population of neutron stars (dotted lines),
  stellar-mass BHs (dashed), and SMBH binaries (solid), as a function
  of orbital radius.}
\begin{center}
\scalebox{0.55}{\includegraphics{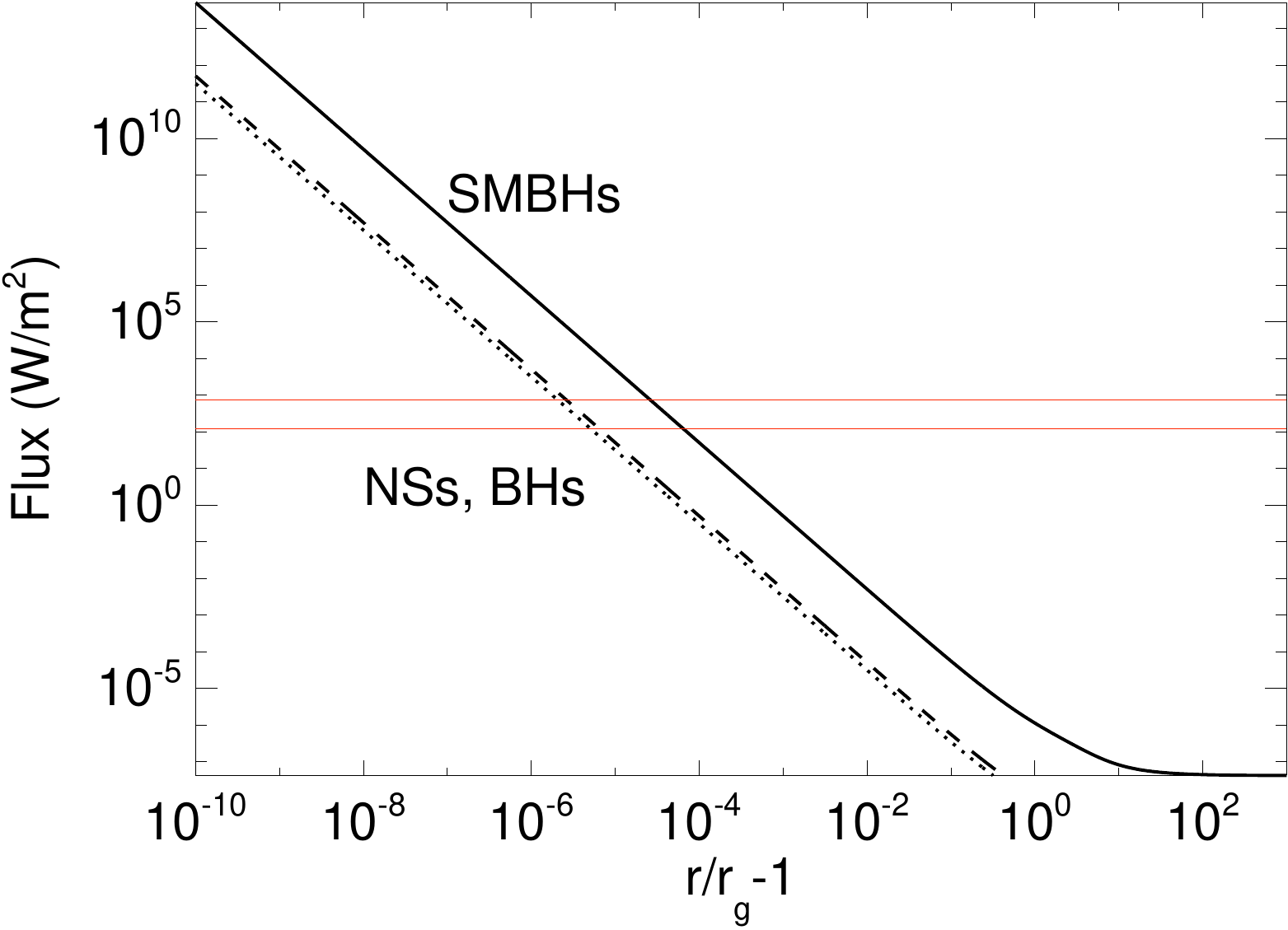}}
\end{center}
\end{figure}

\begin{figure}[h]
\caption{\label{fig:GWfreq} Flux on a planet due to gravitational
  waves from a stochastic population of neutron stars (dotted lines),
  stellar-mass BHs (dashed), and SMBH binaries (solid), as a function
  of orbital radius.}
\begin{center}
\scalebox{0.55}{\includegraphics{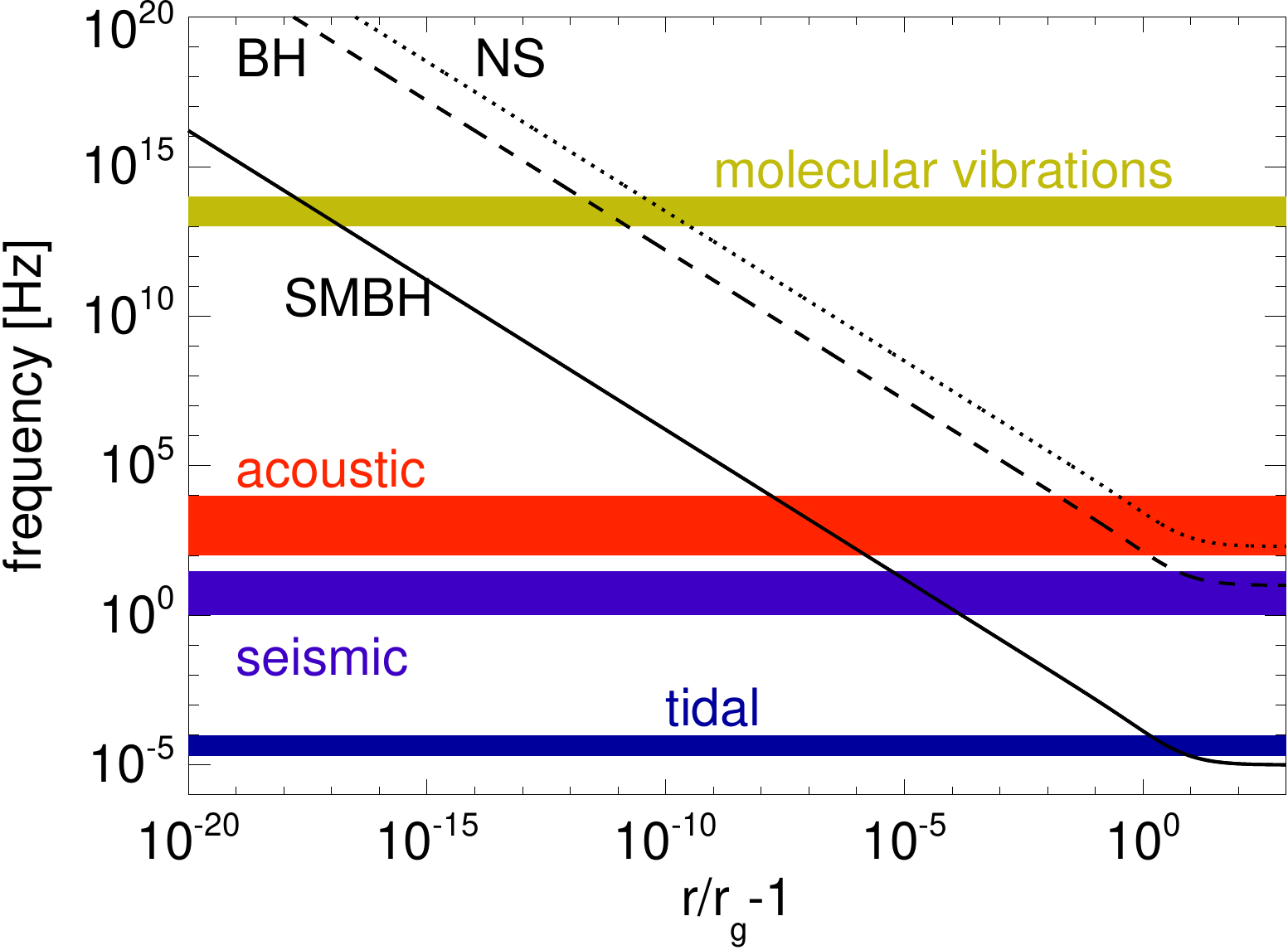}}
\end{center}
\end{figure}

To date (halfway through LIGO's O3 observing run), over thirty BH-BH
mergers have been detected, as well as a handful of NS-NS events. From
these initial detections, we can estimate the event rates in the
local(ish) Universe to be 30 Gpc$^{-3}$ yr$^{-1}$ for the fiducial
30+30$M_\odot$ systems like GW150914, and 600 Gpc$^{-3}$ yr$^{-1}$ for
NS-NS mergers like GW170817 \citep{LSC2018}. For simplicity, we
further assume that all the energy is released at a single frequency:
5\% of the rest mass for BH-BH, at 150 Hz, and 1\% of the rest mass
for NS-NS, at 3 kHz. This leads to a steady-state flux of GWs
throughout the universe of roughly $6\times 10^{-12}$ W m$^{-2}$.

We may as well add to this the flux from SMBH binary mergers, whose
rates are not well constrained, but one reasonable estimate is that
each Milky Way-type galaxy undergoes one major merger during its
lifetime. The SMBH density at $z=0$ is roughly $10^6 M_\odot$
Mpc$^{-3}$ \citep{Li2011}, dominated by black holes in the
$10^{7-8}M_\odot$ range. This gives a merger rate of $10^{-3}$
Gpc$^{-3}$ yr$^{-1}$, or a flux of $10^{-8}$ W m$^{-2}$,
peaked around 0.045 mHz. The flux in a gravitational wave is
proportional to $\dot{h}^2$, the derivative of the strain amplitude
squared, so the observed flux scales like the GW frequency
squared. Thus the same blueshifting effects so important to
electromagnetic radiation will also amplify the GW flux seen by a
planet near the black hole horizon.

In Figure \ref{fig:GWflux} we plot the GW flux due to stellar-mass BHs
and NSs, as well as SMBHs, which turn out to dominate the
background. As before, we mark the narrow band of habitability,
corresponding to $r \approx 1.00003 r_g$, right around the location of
Miller's planet! However, we have not discussed how this GW power
might actually couple to the planet to provide a viable heating
source. Gravitational waves are notoriously bad at coupling to matter,
which is why it took a century between their prediction and
detection. 

In Figure \ref{fig:GWfreq}, we consider a few conceivable physical
coupling mechanisms between the GW background and the planet. Again,
note the expanded x-axis, required to cover the huge dynamic range of
physical processes under consideration. At the
lowest frequencies are tidal forces, operating on the timescales of
hours (just like in Miller's planet! \citet{Thorne:2014}). At somewhat
higher frequencies, global seismic modes (1-10 Hz) or mechanical
accoustic vibrations (100-2000 Hz) become important. Here, the GWs
could make the planet hum quietly, but still lack the necessary power
to do any significant work. At much higher frequencies, we reach the
band of molecular vibrations, coinciding with infrared absorption
bands in the $10^{13-14}$ Hz range. Ironically, this brings us
full-circle to our initial discussion of habitability and energy
absorption in Earth-like atmospheres. Except now, the molecules would
be excited by gravitational, not electromagnetic radiation.

Like EM radiation, we imagine the coupling will be most efficient near
resonant transitions. For argument sake, we take the typical molecular
vibration bands to centered at $10^{13.5}$ Hz, and treat the molecule
like a mechanical ball-and-spring toy model, physically driven by the
gravitational waves, and with a resonance width of
$\delta\omega/\omega_0 = 10^{-3}$. Then the power delivered to the
resonator is given by a Lorentzian function, just like a classical
damped, driven harmonic oscillator. Again, we consider absorption in
both the atmosphere, and interior of the planet. 

\begin{figure}[h]
\caption{\label{fig:GWpower} Absorbed flux on a planet due to gravitational
  waves from a stochastic population of neutron stars, stellar-mass
  BHs, and SMBH binaries, as a function of orbital radius. The flux is
absorbed via molecular vibrations in the atmosphere (solid curve) or
the planet interior (dashed curve).}
\begin{center}
\scalebox{0.55}{\includegraphics{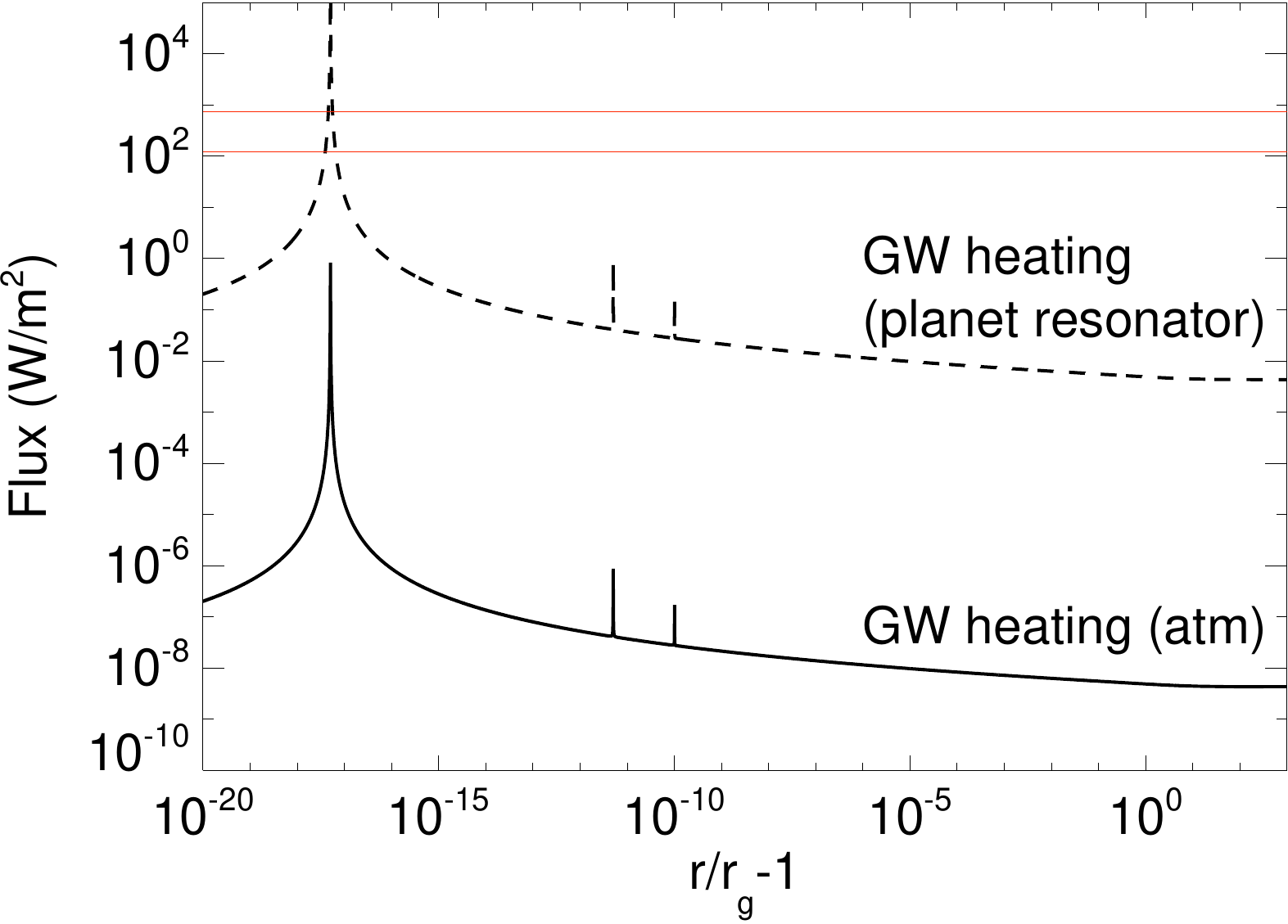}}
\end{center}
\end{figure}

Combining the three
source classes, the absorbed power is plotted in Figure
\ref{fig:GWpower}. We are left with a single, precariously narrow band
of habitability at $r/r_g-1\approx 5\times 10^{-18}$. As with most of
the other energy sources considered in this paper, resonant molecular
heating from GWs would likely be extremely unpleasant to experience,
as the very fabric of our existence would be shaken to its very core.

\section{Discussion}\label{section:discussion}

In this paper, we have explored the possibility of a habitable
Earth-like planet in orbit around a supermassive black hole. At times
tongue-in-cheek, and at times fantastical, we nonetheless found the
study informative, and a useful pedagogical tool for introducing
students (and professionals) to a wide range of fascinating
astrophysical topics.

Just like in the movie {\it Interstellar}, the best science fiction
stories are the ones that can push us to test the limits of our own
scientific understanding, especially in the extreme cases found around
black holes. As we have seen, it is not just the
irreversible nature of the 
black hole's event horizon that can drive the narrative of a science
fiction story, but the bizarre effects that gravity has on time and
trajectories or particles and light. For ultimately, what is
narrative, if not the description of our own passage through time and
space? 



\section*{Acknowledgments}

This work was inspired and motivated by the film {\it Interstellar},
directed by Christopher Nolan and advised by Kip Thorne. We gratefully
acknowledge support from the Goddard Interdisciplinary Science Task
Group program initiated by Piers Sellers. Zachary Shrier provided
invaluable comments and discussion of the film and paper.

\newpage
\bibliography{hzsmbh.bib}

\end{document}